\documentclass[aps,prb,preprint,onecolumn,citeautoscript]{revtex4-1} 
\usepackage{amsmath,amssymb,mathrsfs,bm,feynmf,setspace}
\usepackage{graphicx}
\usepackage[tight]{subfigure}  
\usepackage{color} 
\usepackage[papersize={8.5in,11in}]{geometry}
\usepackage[colorlinks=true]{hyperref} 
\hypersetup{
    bookmarks=true,         % show bookmarks bar?
    unicode=false,          % non-Latin characters 
    pdftoolbar=true,        % show Acrobat
    pdfmenubar=true,        % show Acrobat 
    pdffitwindow=false,     % window fit to page when opened
    pdfstartview={FitH},    % fits the width of the page to the window
    pdftitle={My title},    % title
    pdfauthor={Author},     % author
    pdfsubject={Subject},   % subject of the document
    pdfcreator={Creator},   % creator of the document
    pdfproducer={Producer}, % producer of the document
    pdfkeywords={keyword1} {key2} {key3}, % list of keywords
    pdfnewwindow=true,      % links in new window
    colorlinks=true,       % false: boxed links; true: colored links
    linkcolor=red,          % color of internal links (change box color with linkbordercolor)
    citecolor=blue,        % color of links to bibliography
    filecolor=magenta,      % color of file links
    urlcolor=cyan           % color of external links
} 
\newcommand{\al}{\alpha} 
\newcommand{\be}{\beta} 
\newcommand{\g}{\gamma}
\newcommand{\de}{\delta} 
\newcommand{\e}{\epsilon} 
 
\newcommand{\z}{\zeta}

\newcommand{\ka}{\kappa}
\newcommand{\la}{\lambda}

\newcommand{\s}{\sigma}

\newcommand{\w}{\omega}

\newcommand{\pd}{\partial}

\newcommand{\ang}[1]{\left\langle #1 \right\rangle}

\newcommand{\beq}{\begin{equation}}
\newcommand{\eeq}{\end{equation}}
\newcommand{\Beq}{\begin{eqnarray}}
\newcommand{\Eeq}{\end{eqnarray}}
\newcommand{\bml}{\begin{multline}}

\newcommand{\eeqm}{\end{multline}}

\newcommand{\bsp}{\begin{split}}
\newcommand{\esp}{\end{split}}

\renewcommand{\b}[1]{{\bm #1}}

\newcommand{\inv}{^{-1}}

\newcommand{\mc}{\mathcal}

\newcommand{\ra}{\rightarrow}
\newcommand{\req}[1]{Eq.~(\ref{eq:#1})}

\newcommand{\rfig}[1]{Fig.~\ref{fig:#1}}
\newcommand{\nn}{\nonumber}

\DeclareMathOperator{\diag}{diag}
\DeclareMathOperator{\sgn}{sgn}

\begin{document}

\title{Dispersing quasinormal modes in\\ 2+1 dimensional conformal field theories}
 \author{William Witczak-Krempa}
 \affiliation{Perimeter Institute for Theoretical Physics, Waterloo, Ontario N2L 2Y5, Canada}
 \author{Subir Sachdev}
 \affiliation{Department of Physics, Harvard University, Cambridge, Massachusetts, 02138, USA}
 \date{\today\\
 \vspace{1.6in}}
\begin{abstract}
We study the charge response of conformal field theories (CFTs) 
at non-zero temperature in 2+1 dimensions using the AdS/CFT correspondence.
A central role is played by the quasinormal modes (QNMs), specifically,
the poles and zeros of the current correlators. We generalize our recent study of the QNMs of the a.c.\ 
charge conductivity to include momentum dependence. This sheds light on the various excitations
in the CFT.
We begin by discussing the R-current correlators of the $\mc N=8$ SU$(N_c)$ super-Yang-Mills theory at its
conformal fixed point using holography.
For instance, transitions in the QNM spectrum as a function of momentum clearly identify 
``hydrodynamic-to-relativistic''crossovers. We then extend our study to include 
four-derivative terms in the gravitational description allowing us to study more generic charge
response as well as the role of S-duality, which plays a central role in understanding the
correlators. The presence of special purely-damped QNMs can lead to new behavior, distinct from 
what occurs in the aforementioned gauge theory.
We also extend previous conductivity sum rules to finite momentum and discuss their interpretation in the gravity picture.  
A comparison is finally made with the conformal fixed point of the vector O($N$) model in the large-$N$ limit. 
\end{abstract}
\maketitle
\tableofcontents 

\section{Introduction}
\label{sec:intro}

The subject of strongly-coupled quantum criticality (QC) without well-defined quasiparticle excitations has long been an important
focus of the study of correlated electron systems. The simplest examples of such states in two spatial dimensions
are provided by quantum critical points described by conformal field theories (CFT3s). The traditional condensed matter
approach to the non-zero temperature dynamics of such systems has been to apply the quantum Boltzmann equation,
and related perturbative field-theoretic methods\cite{book}. However, such methods are designed for systems with long-lived
quasiparticles, and the range of their applicability to systems without quasiparticles is unclear. The AdS/CFT correspondence\cite{Maldacena} 
has provided a new set of tools to investigate the dynamics of QC\cite{m2cft}: an important advantage
of this method is that quasiparticles do not appear at any stage in the computation, and even the leading results do not contain
artifacts linked to the existence of long-lived quasiparticles. Instead, a simple picture of the dynamics emerges in terms of the
{\em quasinormal modes\/} (QNMs) of a gravitational theory on an asymptotically Anti de Sitter (AdS) spacetime. 
In a recent paper we described the structure of these QNMs
in some detail for spatially uniform probes \cite{ws}. This paper will extend the analysis to allow for spatial dependence, and describe
the dispersion of the QNMs as a function of spatial momentum. As we shall see, this yields far deeper insights 
than the zero-momentum response.

Our objects of study are two-point correlation functions of currents in thermally excited CFT3s.
Expressed in frequency-momentum space these are $\langle J_\mu(\w,\b k)J_\nu(-\w,-\b k) \rangle$,
where the average is over the thermal density matrix, while $\mu,\nu$ are spacetime indices. Such
correlators yield, for instance, the 
frequency-dependent charge conductivity in the limit of vanishing momentum. We emphasize that such
current correlators can be probed experimentally in
QC systems with emergent Lorentz invariance. An important
example is the quantum phase transition between a bosonic Mott insulator and a superfluid in two spatial 
dimensions, which is described by the well-known QC O(2) model. 
Recent experiments with ultra-cold atoms in optical lattices
have realized such a QC point, both in three\cite{greiner3D} and two\cite{spielman,endres,zhang} 
dimensions. The excellent control in these experiments gives hope that the universal QC charge 
response could be measured in the near future.
On the other hand, as interactions are strongly relevant in such critical systems, the theoretical description
of many quantities remains very challenging, especially concerning the current correlators of the U(1) charge
in real time and at finite temperature.
Holographic methods have successfully yielded some fresh insights into the problem at 
hand\cite{m2cft,ritz,myers11,ws}.  
For instance, the frequency-dependent conductivity of a strongly correlated CFT (with supersymmetry) has been exactly
computed\cite{m2cft} using the AdS/CFT correspondence\cite{Maldacena}. 
It was remarkably found that the conductivity does not vary with
frequency because of an emergent self-duality, which is not expected to hold for generic CFTs.
Further extensions\cite{ritz,myers11} of the original holographic model have yielded 
frequency-dependent conductivities that can be expected in quantum fluids with particle- or vortex-like
excitations. A sharp distinction between the two types of response can be made by examining the correlators
at complex frequencies\cite{ws}: the particle-like conductivity has a pole at a frequency of order $-iT$, whereas
the vortex-like conductivity has a \emph{zero} there. These become manifest at real frequencies by the presence of 
a maximum or minimum at zero frequency, respectively. 
In the former case, a small-frequency peak is obtained to which we refer as the Damle-Sachdev (DS) 
peak\cite{damle}. At a formal level, it shares the same single-pole structure as the 
Drude peak characterizing the transport of a gas of electrons subject to a source
of inelastic scattering (disorder). An important difference is that 
translation symmetry is not broken in a CFT, instead the finite d.c.\ conductivity results from
the particle-hole symmetry at zero charge density. This is why we choose the designation DS instead of ``Drude''.
\footnote{We thank one referee for making this suggestion.}

The poles and zeros of the retarded correlators, the QNMs, acquire a significant physical meaning for the 
correlated CFT: they are excitation modes of the system. In this sense, understanding how they propagate or
disperse as a function of momentum yields important insight into the excitation channels of the CFT. 
In a certain sense, the QNMs supersede the concept of weakly interacting quasiparticles. 
Interestingly, the AdS/CFT correspondence connects these QNMs with discrete excitation 
modes of a black hole in one higher
dimension\cite{star1,nunez,star2}, which allows for the inclusion of a finite temperature.
In the case of 
the QNMs of the current correlators, these excitations correspond to electromagnetic ``eigenmodes'' of the black hole.
The holographic correspondence thus identifies the ``normal'' modes of black holes with the emergent
excitations of a strongly correlated CFT\cite{star1,nunez,star2}. We mention that such QNMs have been explored 
in a variety of different holographic applications to strongly correlated systems
potentially relevant to condensed matter\cite{m2cft,hh,hSC,denef,bhaseen,sum-rules,ws}. 
A recent review by Hartnoll\cite{sean-rev} covers a number of these, with special focus on the QNMs.

\subsection{Main results}
\label{sec:main}
The CFT current correlators are obtained holographically via a general four-derivative
bulk action\cite{ritz,myers11}, which in one limit describes the R-current correlators 
of an $\mc N=8$ superconformal
gauge theory\cite{m2cft}. The correlators are found to bear a strong imprint
from their $T=0$ Lorentz invariant form, such as a ``reflection'' property between the real and imaginary
parts under exchange of frequency and momentum, illustrated in \rfig{C}. 
In almost the entire frequency-momentum domain, except 
in the hydrodynamic regime, the correlators can indeed be interpreted as ``smoothed'' versions of the relativistic
forms. The ``smoothing'' occurs via the breakup at finite temperature of the $T=0$ branch cuts into a 
discrete sequence of poles and zeros, the QNMs. 
We further identify sharp transitions in the QNM spectrum as a function of momentum, and these
manifest themselves at real frequencies as hydrodynamic-to-relativistic crossovers. In this respect, we emphasize how
the presence of a four-derivative term in the bulk action can lead to distinct behavior
compared to the leading two-derivative Einstein-Maxwell action. The latter is self-dual under an electric-magnetic
duality\cite{m2cft}, and this constrains the correlators considerably. Higher order terms will break this
self-duality\cite{myers11}, and in particular give rise
to a special QNM on the imaginary frequency axis, referred to as the D-QNM due to its 
purely damped nature and its formal relation to the usual Drude scaling. This QNM lies at the heart 
of the difference between the particle- and vortex-like responses and it 
is expected to be present in generic CFT3s\cite{ws,will},
and indeed appears in the four-derivative holographic theory. As we have noted
above, when this D-QNM is a pole, it gives rise to a small-frequency DS peak in the conductivity, 
whereas a valley results when it is a zero.

A generalization of particle-vortex duality, S-duality\cite{witten}, which manifests itself as 
electric-magnetic duality in the bulk, plays a central role in our analysis. It leads for instance to the
appearance of a hydrodynamic zero in the transverse response (i.e.\ when the current is transverse to the momentum).
This zero becomes the standard hydrodynamic pole $\w \sim -i\hat D k^2$ of the S-dual theory, where $\hat D$ is the
S-dual diffusion constant. 
S-duality is also central in the sum rule analysis, which we extend to finite momentum, see \req{sr1} and 
\req{sr2}. The conductivity sum rules\cite{sum-rules,ws} obtained at zero momentum, 
for both the direct and S-dual\cite{ws} theories, are found to rely on the bulk gauge invariance of the 
gauge field holographically dual to the CFT current operator. 
We finally conclude with a comparison with the vector O($N$) model at its large-$N$ conformal fixed point. For instance,
we find an analogous sum rule to what is obtained in the holographic analysis, including an almost exact agreement between
static correlators entering the sum rules. 

The outline is as follows: We first discuss general properties of current correlators in CFT3s in Section~\ref{sec:cft}.
We then turn to their explicit study in a supersymmetric gauge
theory using the AdS/CFT correspondence in Section~\ref{sec:sym}. We extend the analysis to more generic
holographic models including a four-derivative bulk term in Section~\ref{sec:gamma}. Sum rules are discussed in
Section~\ref{sec:sr}, and finally a comparison with the vector O($N$) model is made in Section~\ref{sec:on},
followed by a conclusion. 

\section{Current correlators in a CFT}
\label{sec:cft}
We first review some general properties of the current correlators in CFT3s, namely
the meromorphic structure in Subsection~\ref{ssec:mero} and the asymptotics in Subsection~\ref{ssec:asym}.

\subsection{Physical imprint in the analytic structure}
\label{ssec:mero}
In the holographic models we consider below, the finite-temperature current correlators will be meromorphic in the 
complex frequency plane, i.e.\ they will be analytic except at a discrete set of
finite order poles. Moreover all the poles (and zeros) will be in the lower half-plane (LHP) of
frequency by virtue of retardedness. 
One can ask: to what extent are these properties generic? For instance, it is well-known that the zero-momentum 
correlators, which give the conductivity, will generically have a branch cut at frequencies 
whose norm is much less than the temperature. This branch cut emanates from the zero frequency point
and is associated with the so-called long-time tails of hydrodynamics\cite{kovtun-ltt,kovtun-rev}, following 
from the presence of gapless hydrodynamic modes at arbitrary long-wavelengths, such as the well-known 
diffusive mode discussed below. 
At finite momentum, however, the length scale introduced in the problem is expected to introduce
an IR cutoff beyond which correlators decay exponentially and not algebraically, thus excluding branch
cuts of that sort. Such arguments seem reasonable as they are based on the universal principles of hydrodynamics.
But what about frequencies whose norm is of the order or greater than the temperature and thus fall
beyond the hydrodynamic regime? In that case,
a statement regarding general interacting CFTs is hard to establish, but it is not unreasonable to
expect that a slightly (read infinitesimally) 
perturbed thermal CFT state will relax back to {\em local\/} equilibrium exponentially fast, which is tantamount
to assuming that no branch cuts will emanate from the real axis. This does not preclude the 
presence of branch cuts in the LHP away from the real frequency-axis. These are absent
in the holographic models we study, but it is at present unclear if CFT3s will generically 
obey this rule. It would be interesting to investigate this aspect in more detail by
considering specific CFTs, such as the O$(N)$ vector model at finite but large $N$. 
(The $N\ra\infty$ is discussed in Section~\ref{sec:on}.) 

To gain further insight into the role of the QNMs, let us consider a generic meromorphic
current correlator, as obtained using holography for instance:
\begin{align}\label{eq:mero}
  \langle J_\mu(\w,\b k)J_\nu(-\w,-\b k) \rangle &=a \frac{\prod_m (\w-\hat \w^m_k)}{\prod_n (\w-\w^n_k)} \nn\\
  &\ra \sum_n a_n \exp\left(-|\Im \w_k^n| t -i \Re \w_k^n t\right)\,,
\end{align}
where $a$ is a constant and the arrow indicates a temporal Fourier transform. Note that 
we are leaving out the 
delta function $\ang{ J_\mu(k_\al)J_\nu(k_\al')}=\de^{(3)}(k_\al+k_\al')C_{\mu\nu}(k_\al)$. 
Each pole QNM, $\w_k^n$, contributes an
exponential to the time-dependence of the correlation function. As all the poles
are in the LHP, the exponentials decay in time. 
%(Poles directly on the real-frequency axis lead to power law decay instead.)
We thus see that the QNMs closest to the real axis will dominate the long-time response 
of the system since they correspond to the longest decay time-scales. 
On the other hand, the real part of the QNM frequency provides the oscillatory behavior. 
As the momentum $k$ increases from zero to infinity, these time scales will change and lead to very different
behavior. In the language of the excitation modes of the system, the absolute value of the imaginary part
gives the lifetime of the excitation, while the real part its energy. A so-called quasiparticle mode
will have an energy much greater than its lifetime; while purely damped modes do not have a real part, such
as the diffusion pole. 
The momentum dependence of these modes gives generalized dispersion relations for the excitations. In the case of the
quasiparticle QNMs, this replaces the usual energy-momentum dispersion relation of weakly interacting quasiparticles.

The zeros, $\hat\w_k^n$, play an important role as well. Not only are they essential to determine the
values of the correlator, but in the theories we consider below,
they also correspond to the QNM excitations of the S-dual theory. S-duality (see \onlinecite{witten}
for its action on CFT3s) is a generalization of 
the usual particle-vortex duality familiar in the context of the O(2) model describing the superfluid-to-insulator
quantum phase transition.

Finally, as we argue below, the QNMs evolve into branch cuts in the limit of zero temperature
where one recovers power law decay in time of the correlation functions, in accordance with the behavior 
expected to take place directly at the fixed point. 
This is physically clear given that the QNM excitations merge and form a continuum.

\subsection{Symmetries and asymptotic forms}\label{ssec:asym}
We now review the symmetries and asymptotic behavior of the current correlators. 
This will serve as an important comparison point throughout the work.
Although most of the discussion follows Ref.~\onlinecite{m2cft}, we make new remarks regarding the
hydrodynamic behavior of the transverse current correlators, which naturally leads to S-duality. 
We consider correlation functions involving two conserved currents, $J_\mu^a(x)$, 
in a CFT at finite temperature, where $\mu$ is the spacetime index, while $a$ labels the flavor. 
%We review the general properties of these correlators, which 
%include the presence of diffusive poles and \emph{zeros} in the hydrodynamic limit.
The Fourier transform of the retarded
current correlator, $C_{\mu\nu}^{ab}(x)=-i\theta(x^0)\langle [J_\mu^a(x),J_\nu(0)] \rangle$, can be decomposed as follows
\begin{align}
  C_{\mu\nu}^{ab}(\w,\b k)=P_{\mu\nu}^L\Pi^L_{ab}(\w,k) + P_{\mu\nu}^T\Pi^T_{ab}(\w,k)\,,
\end{align}
where two sets of functions, $\Pi_{ab}^{L}$ and $\Pi_{ab}^{T}$, are needed because the temperature breaks
the Lorentz invariance. Note that $C_{\mu\nu}^{ab}/T$
depends only on the rescaled frequency $\w/T$
and momentum $\b k/T$, the latter ratio arising since a CFT has a dynamical exponent $z=1$. 
We work in units where $c=\hbar=k_B=1$ throughout.
The transverse projector reads%, $P^{L,T}_{\mu\nu}$, read respectively: 
\begin{align}
  P_{tt}^T=P_{ti}^T=P_{it}^T=0, \quad P_{ij}^T=\de_{ij}-\frac{k_ik_j}{\b k\cdot\b k}\,,
\end{align}
and by orthogonality: $P_{\mu\nu}^L=[\eta_{\mu\nu}-k_\mu k_\nu/k\cdot k]-P_{\mu\nu}^T$,
where $k^\mu=(\w,\b k)$ and roman indices run over spatial coordinates. The Minkowski
metric was introduced, $\eta_{\mu\nu}=\diag(-1,1,1)$, such that $k\cdot k=\eta_{\la\la'}k^{\la}k^{\la'}=-\w^2+k^2$.
It is easy to see that $C_{\mu\nu}^{ab}(\w,\b k)$ is symmetric in $\mu,\nu$ and satisfies the Ward identity 
$k^\mu C_{\mu\nu}^{ab}(\w,\b k)=0$ resulting from current conservation. Both of these properties are in fact 
independently satisfied by the projectors, $P_{\mu\nu}^{L,T}$.

Due to the rotational invariance, we are free to fix the momentum to point along the $x$-direction, $\b k=(k,0)$, which yields
\begin{align}
  C_{tt}^{ab}(\w,\b k) &=\frac{k^2}{\w^2-k^2}\Pi^L_{ab}\,; \quad \label{eq:ctt}
  C_{xx}^{ab}(\w,\b k) =\frac{\w^2}{\w^2-k^2}\Pi^L_{ab}\,, \\
  C_{yy}^{ab}(\w,\b k) &=\Pi^T_{ab}\,. \label{eq:cyy}
\end{align}
Also, $C_{tx}^{ab}=C_{xt}^{ab}=-\frac{\w k}{\w^2-k^2}\Pi^L$ while all the ones
mixing $y$ with $x$ or $t$ vanish, in line with the decoupling between longitudinal
and transverse responses.
In the limit of zero temperature, a simple form dictated by Lorentz
invariance emerges:
\begin{align}\label{eq:CT0}
  C_{\mu\nu}^{ab}(\w,\b k)\big|_{T=0}= \left(\eta_{\mu\nu}-\frac{k_\mu k_\nu}{k\cdot k}\right)\sqrt{k\cdot k}K_{ab}\,,
\end{align}
where $K_{ab}$ are the conductivities in the $\w/T\ra\infty$ limit: $K_{ab}=\s^{ab}_\infty$.
We choose the branch of the square root $\sqrt{k\cdot k}$ to be in the LHP $\w$-plane 
and such that $\Im\sqrt{k\cdot k}>0$ when $w>q$.
Specifically, the charge and transverse current correlators
read: 
\begin{align}
  C_{tt}^{ab}(\w,\b k)\big|_{T=0} &=\frac{k^2}{\sqrt{-\w^2+k^2}}K_{ab}\,,\\
  C_{yy}^{ab}(\w,\b k)\big|_{T=0} &=-\sqrt{-\w^2+k^2}K_{ab}\,.
\end{align}
We note that these functions have branch points at $\w=\pm k$: $C_{tt}$ has branch poles while
$C_{yy}$ branch zeros. 
% The presence of branch cuts implies the absence of undamped, 
% propagating modes, which would be signaled by simple poles at $\w=\pm k$.

At finite temperature and in the opposite limit of small frequency and momentum, $|\w|,k\ll T$, we obtain hydrodynamic behavior. 
For example, the correlators of the conserved
charge densities, $J_t^a$, read
\begin{align}
  C_{tt}^{ab}(\w,k) &\approx \sum_\ell  \frac{\chi_{ab}^{\ell} D_{\ell} k^2}{i\w-D_\ell k^2}\,,  \quad |\w|,k\ll T\,,
\end{align}
where $\ell$ sums over the diffusive eigenmodes; $\chi_{ab}^\ell,\, D_\ell$ are the corresponding charge 
susceptibilities and diffusion constants, respectively. 
By virtue of scaling: 
$\chi_{ab}=T \bar\chi_{ab}^\ell$ and $D_\ell=\bar D_\ell/T$, with $\bar\chi_{ab}^\ell,\bar D_\ell$
being universal dimensionless quantities associated with the conformal fixed point, just like the
$K_{ab}$ introduced in \req{CT0}. They are related to the d.c.\ conductivities via Einstein relations:
$\s^{ab}_{\rm dc}=\sum_\ell \chi_{ab}^\ell D_\ell$. Crucially, the hydrodynamic charge response is characterized by the presence of 
diffusive poles in the LHP $\w$-plane at $\w=-iD_\ell k^2$.
In the same limit, we propose that the transverse current correlator is given by
\begin{align}\label{eq:Cyy-hydro}
  C_{yy}^{ab}(\w,k) &\approx -\sum_\ell \chi_{ab}^{\ell}  D_{\ell}(i\w-\tilde D_\ell k^2)\,,
\end{align}
where the tilde variables $\tilde D_\ell$ obey the same scaling as their cousins $D_\ell$. We shall see that in theories
for which we can define S-duality, these are the diffusion constants of the S-dual theory. 
Contrary to $C_{tt}$, the hydrodynamic behavior of the transverse correlator is analytic.
However, it has analogs to the diffusive poles: a set of ``diffusive zeros'' at $\w=-i \tilde D_\ell k^2$.
When S-duality exists, it will map these to the diffusive poles of the S-dual theory. 

\section{$\mc N=8$ super-Yang-Mills}
\label{sec:sym}
We first examine a special theory whose holographic description is believed to be very simple:
it is a 2+1D Yang-Mills theory with gauge group SU$(N_c)$ and $\mc N=8$ supersymmetry\cite{abjm}. In 
a certain large-$N_c$ limit, the theory flows to a strongly coupled CFT. Further, it is believed that
this CFT admits a holographic description in terms of a string theory (or rather its
10+1D extension, M-theory). The holographic duality maps the vacuum of the CFT to a stack of $N_c$ M2-branes which can
be described by M-theory on AdS$_4\times S_7$. In the large-$N_c$ limit, the M-theory
reduces to classical supergravity and the AdS/CFT correspondence allows one to
relate the correlators of the classical gravity on AdS$_4$ to those of the CFT 
living in one lesser spatial dimension. In particular, the CFT has a set of SO$(8)$
R-symmetries, which map to the symmetries of the sphere $S_7$ via the holographic
correspondence. The R-symmetries can be thought of as rotations amongst the $\mc N=8$ supercharges,
which get mapped to rotations of the $7$-sphere in the dual theory.

The correlators of the $28={8\choose 2}$ R-currents $\{J_\mu^{a=1,\dots,28}\}$ are entirely encoded
in two functions, $\Pi^{L,T}$, due to the SO$(8)$ symmetry:
\begin{align}
  C_{\mu\nu}^{ab}(\w,\b k)=\de_{ab}\left[P_{\mu\nu}^L\Pi^L(\w,k)+P_{\mu\nu}^T\Pi^T(\w,k) \right]\,.
\end{align}
Only the diagonal correlators remain finite allowing us to drop the flavor indices. 
A further simplification was shown to exist\cite{m2cft} because
of the presence of a self-duality which manifests itself as an electric-magnetic duality in the
gravitational description. As a result, the longitudinal and transverse correlators are directly related:
\begin{align}\label{eq:scft_duality-rel}
  \Pi^L(w,q)\Pi^T(w,q)=\chi_0^2(-w^2+q^2)\,,
\end{align}
where we have introduced the rescaled frequency and momentum:
\begin{align}
  w=\frac{3\w}{4\pi T}\,, \qquad   q=\frac{3k}{4\pi T}\,. \label{eq:rescaling}
\end{align}
We have also introduced the charge susceptibility\cite{m2cft} $\chi_0=(4\pi T/3) g_4^{-2}$, where $g_4^{-2}=(\sqrt 2/6\pi)N_c^{3/2}$
is the inverse coupling squared of the gauge field holographically dual to a given R-current, as discussed
in more detail below.
It should be noted that that the frequency and momentum are rescaled by the diffusion constant, $D_0=3/4\pi T$,
which is related to $\chi_0$ by the Einstein relation $D_0\chi_0=\s_0$, where the d.c.\ R-charge conductivity
is $\s_0=1/g_4^2$. By virtue of the self-duality
mentioned above, the a.c.\ conductivity was remarkably found to be frequency independent\cite{m2cft}, a fact
to which we return in the next section. 

We finally note that as a result of \req{scft_duality-rel}, the charge correlator $C_{tt}$, \req{ctt}, can be expressed in terms
of the transverse one, $\Pi^T=C_{yy}$:  
\begin{align}
  C_{tt}(w,q)=-\frac{\chi_0^2 q^2}{\Pi^T(w,q)}
\end{align}

\subsection{Bulk action for boundary correlators}
We introduce the basic holographic tools needed to compute the correlators and refer the reader to
some reviews aimed at condensed matter researchers for further background on this rich 
topic\cite{sean-rev,mcgreevy-rev,ss-rev,sachdev-rev}.
In the AdS/CFT correspondence, the global currents in the CFT are dual to gauge fields in the bulk.
Since the current correlators are diagonal in the R-charge flavor index, and the bulk gauge coupling
tends to zero in the large-$N_c$ limit we can focus on a
single $U(1)$ gauge field $A_a(t,x,y,r)$ instead of considering the full non-abelian
SO(8) gauge structure. The coordinate $r$ is along the extra spatial dimension. 
This gauge field will be dual to a current operator $J_\mu(t,x,y)$ that can be thought
to live on the boundary at $r=\infty$. 
The AdS/CFT correspondence relates
the current correlator to the value of the gauge field at the boundary of the bulk 3+1D spacetime.
One then trades the problem of computing two-point functions in a correlated CFT with
that of solving Maxwell equations for $A_\mu$ in a specific curved spacetime. The latter contains
a (planar) black hole and the $r$-coordinate of the horizon is proportional to the temperature of the boundary CFT.
The spacetime tends to AdS$_4$ as $r$ approaches infinity. The gravitational 3+1D Maxwell-Einstein action
used to calculate the current correlators of the boundary CFT reads\cite{m2cft}
\begin{align}\label{eq:S_bulk}
  S_{\rm bulk}=\int d^4x \sqrt{-g}\left[\frac{1}{2\ka^2}\left(R+\frac{6}{L^2}\right)
  -\frac{1}{4g_4^2}F_{ab}F^{ab}\right]\,, %+\g \frac{L^2}{g_4^2}C_{abcd}F^{ab}F^{cd}\right]\,,
\end{align}
where $g$ is the determinant of the metric $g_{ab}$ with Ricci scalar $R$; $F^{ab}$ is the field strength tensor
of the probe U(1) gauge field $A_a$, where roman indices run over the bulk spacetime components, $(t,x,y,r)$.
$L$ is the radius of curvature of the AdS$_4$ spacetime while the gravitational constant $\kappa^2$ is related to 
the coefficient of the two-point correlator of the stress-energy tensor $T_{\mu\nu}$ of the 
boundary CFT (this is reviewed in Ref.~\onlinecite{suvrat} for e.g.), an analog of the central charge of CFTs in 1+1D.
The gauge coupling constant $g_4^2 = 1/\sigma_\infty$ dictates the infinite-$w$ conductivity.

In the absence of the gauge field, which is here only a probe field used to calculate the linear response, 
the metric that solves the gravitational EoM associated with $S_{\rm bulk}$ is:
\begin{align}\label{eq:metric-r}
  ds^2=\frac{r^2}{L^2}\left(-f(r)dt^2+dx^2+dy^2\right) + \frac{L^2dr^2}{r^2f(r)}\,,
\end{align}
where $f(r)=1-r_0^3/r^3$. Such a metric describes a spacetime with a planar black hole whose event horizon 
is located at $r=r_0$ and singularity at $r=0$, and that asymptotically tends to AdS$_4$ as $r\rightarrow\infty$. 
The position of the event horizon is directly proportional to the temperature of the boundary CFT,
\begin{align}
  T=\frac{3r_0}{4\pi L^2}\,.
\end{align}
As $T\rightarrow 0$, the black hole disappears and we are left with pure AdS$_4$, which is
holographically dual to the vacuum of the CFT. The presence of a horizon permits the 
study of thermal states since the energy that is Hawking radiated from it ``heats up the boundary''.
It will be more convenient to use the dimensionless coordinate $u=r_0/r$, in terms of which \req{metric-r} becomes
\begin{align}\label{eq:metric}
  ds^2=\frac{r_0^2}{L^2u^2}\left(-f(u)dt^2+dx^2+dy^2\right) + \frac{L^2du^2}{u^2f(u)}\,, \qquad f(u)=1-u^3\,.
\end{align}
The EoM for the probe gauge field is then the Maxwell equation
$\nabla_aF^{ab}=0$,
where $\nabla_a$ denotes a covariant derivative with respect to the background metric, $g_{ab}$. As we are interested in the
current correlator in frequency-momentum space, we Fourier transform the gauge field:
\begin{align}
  A_a(t,x,y,u)=\int \frac{d^3k}{(2\pi)^3} e^{-i\w t+i\b k\cdot\b x}A_a(\w,k_x,k_y,u)\,,
\end{align}
where the coordinate $u$ was left un-transformed since there is no translational invariance
in that direction. We shall actually solve for the full $u$-dependence of $A_a$. We work in the
radial gauge $A_u=0$. Without loss of generality, we also set the spatial momentum to be along the
$x$-direction, $(k_x,k_y)=(k,0)$. 
As a result of the self-duality described above, to obtain the full charge response
we only need to solve for the transverse correlator $C_{yy}=\Pi^T$.
It can be obtained using the AdS/CFT dictionary:
\begin{align}\label{eq:dict-PiT}
  \Pi^T(w,q) = \left. -\chi_0\frac{\pd_u A_y}{A_y}\right|_{u=0}\,,
\end{align}
where $A_y$ is the transverse gauge mode with 3-vector $(w,q,0)$.
The explicit EoM for $A_y$ is\cite{m2cft}
\begin{align}\label{eq:eomAy}
  A_y''+\frac{f'}{f}A_y'+\frac{w^2-q^2f}{f^2}A_y=0
\end{align}
where primes denote $u$-derivatives. Note that
we are using the rescaling introduced above for the frequency and momentum, \req{rescaling}.
To obtain the retarded correlator, we apply an in-falling boundary condition for the waves at the horizon and solve the
equation numerically\cite{m2cft,myers11,ws}. 

In the next section, we examine the frequency-momentum dependence of these correlators in detail. 

\subsection{Familiar behavior of the current correlators}
Before looking into the QNMs of the current correlators, which correspond to complex frequencies in the
LHP, we first study their behavior at real frequencies.
This has been previously done using the AdS/CFT correspondence in Ref.~\onlinecite{m2cft}. We briefly review the 
results and make some new observations. 

\begin{figure}
\centering%
 \subfigure[~$\Re C_{tt} /q^2$]{\label{fig:ReCtt}\includegraphics[scale=.55]{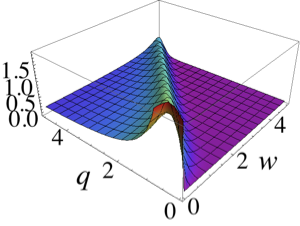}}  
 \subfigure[~$\Im C_{tt}/q^2$]{\label{fig:ImCtt} \includegraphics[scale=.55]{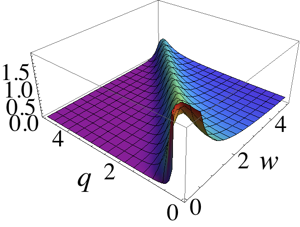}}\\
 \subfigure[~$-\Re C_{yy}$]{\label{fig:ReCyy} \includegraphics[scale=.5]{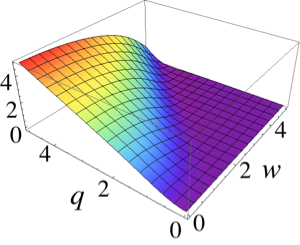}}
 \subfigure[~$\Im C_{yy}$]{\label{fig:ImCyy}\includegraphics[scale=.5]{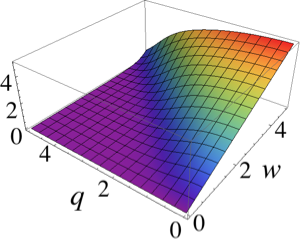}}
\caption{\label{fig:C} Frequency and momentum dependence of the longitudinal
and transverse R-current correlators, $C_{tt}$ and $C_{yy}$, respectively, normalized by $-\chi_0$.
Note the reflection property under exchange of $w$ and $q$, a remnant from the Lorentz invariant $T=0$ form.
} 
\end{figure}

\begin{figure}
\centering%
\includegraphics[scale=.6]{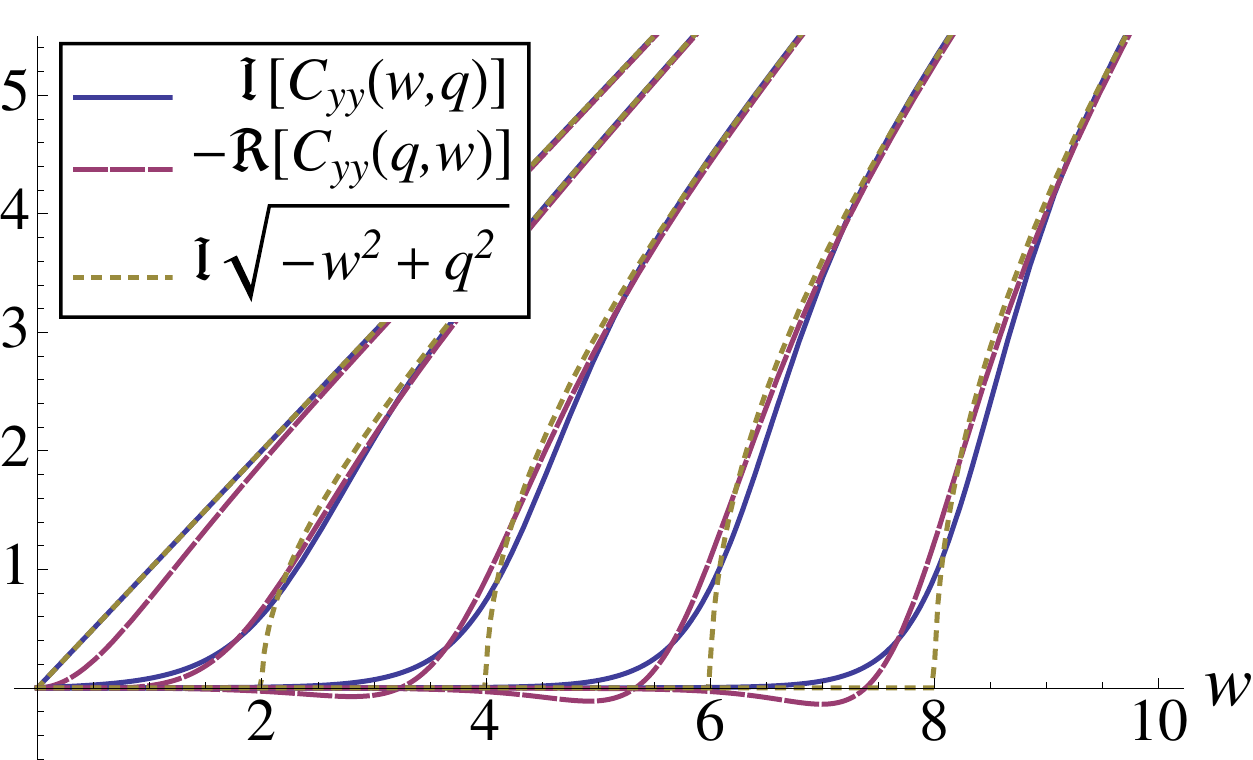}%.48   
\caption{\label{fig:Cyy-vs-w} Comparison of $\Im C_{yy}(w,q)$ with the zero temperature form $\sqrt{-w^2+q^2}$.
From left to right: $q=0,2,4,6,8$. The plots of $C_{yy}$ are in units of $-\chi_0$. 
The plot of $-\Re C_{yy}(q,w)$ (note the arguments are interchanged) illustrates the ``reflection property''.
} 
\end{figure}

The numerical solution can be found in \rfig{C}, where the real and imaginary parts of
$C_{tt}$ and $C_{yy}$ are shown in the $(w,q)$ plane. A salient feature is that the 
real parts of $C_{tt}$ and $C_{yy}$ seem to be mirror images of the imaginary parts with respect to the $w=q$ line.
This is shown more clearly in \rfig{Cyy-vs-w}, where the red dashed lines show the real part reflected along the $w=q$ line, $-\Re C_{yy}(q,w)$.
The agreement is excellent away from the region $w\sim q$. 
% With regard to the latter, %region $w\sim q$,
% we note that the discrepancy between $C_{yy}(w,q)$ and $\sqrt{-w^2+q^2}$ actually \emph{grows} as $w\sim q$ tends
% to infinity, contrary to naive expectations. 
% The difference between the two increases slowly though, as it can be bounded by $\sqrt w$. 
Further, the real parts of both $C_{tt}$ and $C_{yy}$ are finite mainly in the region $q>w$, while
the imaginary parts have support mostly in the complementary region, $w>q$. We note that
the above properties, that are only approximately true here, are \emph{exactly} satisfied by the zero temperature 
Lorentz invariant correlators,
$C_{yy}=\Pi^T=-\s_\infty\sqrt{-\w^2+k^2}$ and $C_{tt}= k^2\s_\infty/\sqrt{-\w^2+k^2}$.  
In fact, the numerical solution for $\Pi^T$
not only closely resembles $\sqrt{-\w^2+k^2}$ but the quantitative agreement is excellent away from the 
region $\w\sim q$, as is shown in \rfig{Cyy-vs-w}. As expected, the Lorentz invariant form is a better match
at frequencies and momenta greater than the temperature: for e.g., the hydrodynamic limit 
clearly deviates from Lorentz invariance, namely $\Pi^T\sim iw-q^2$.
Nonetheless, we note that the zero temperature form
and the finite-$T$ transverse correlator are exactly equal at all
frequencies at zero momentum:\cite{m2cft} $\Pi^T(w,0)=-\chi_0 i w$.
Similarly, the agreement between the R-charge correlator $C_{tt}$ and the $T=0$ form is also good except in the 
region $w\sim q$, where
the latter has a square root divergence while the former has a finite peak. The height of the peak
grows with momentum thus approaching the $T=0$ form.
It should be again noted that it
is not only the large $w\gg q$ region where the agreement is excellent: when $w<q$, $C_{tt}$ decays
very rapidly to zero, the more so as $q$ grows.  

% From these considerations, we see that the holographic results for the
% correlators bear a close similarity to the zero temperature forms, $C_{yy}=\Pi^T\propto \sqrt{-w^2+q^2}$ and
% $C_{tt}\propto 1/\sqrt{-w^2+q^2}$. 
We have seen that the finite temperature R-current
correlators bear a strong imprint from Lorentz invariance.
It is thus not unreasonable to interpret them as ``smoothed'' versions
of the zero temperature relativistic forms. For example they show very similar behavior under the exchange
of $w$ and $q$, and a similar distribution of spectral weight. 
As we will see below, the branch cuts of the $T=0$ forms can be argued
to transform into the infinite sequence of poles and zeros present at finite temperature, the QNMs.

\subsection{Dispersing quasinormal modes}
A deeper insight into the physics encoded in the current correlators can be gained by examining their behavior
in the complex frequency plane. The finite momentum correlators are meromorphic functions
with a discrete set of poles and zeros in the LHP. These correspond to certain QNMs of the black hole
in the dual gravitational description. To some extent they replace the quasiparticle excitations of weakly
interacting theories, which are clearly lacking for the $\mc N=8$ correlated CFT of interest. In the following we
discuss how these QNMs disperse as a function of momentum, and how this can help quantify hydrodynamic-to-relativistic
crossovers. The essential features of the QNM spectrum have been identified in previous works\cite{cardoso1,miranda}.
Our analysis not only corroborates these results but we make further new observations. This
will also serve as a comparison ground for Section~\ref{sec:gamma}, where we study more general holographic actions.
We also mention that Refs.~\onlinecite{davison4d} (3+1D) and \onlinecite{davison3d}, \onlinecite{brattan} (2+1D) 
have studied closely related
phenomena in holographic models for \emph{doped} CFTs at finite temperature, and have made connections between
hydrodynamic-to-collisionless crossovers and transitions in the QNM spectrum.

We recall that the poles and zeros of the transverse correlation function $C_{yy}(w,q)=\Pi^T$ 
determine the entire meromorphic structure of the current correlators
because of the EM duality enforcing $\Pi^L= \chi_0^2(-w^2+q^2)/\Pi^T$. From \req{dict-PiT}, which states that 
$\Pi^T=-\chi_0\pd_u A_y(u=0)/A_y(u=0)$, 
we see that the frequencies
and momenta at which $A_y(u=0)$ vanishes correspond to the poles of $\Pi^T(w,q)$.
Equivalently, the zeros of $\pd_u A_y(u=0)$ give the zeros of $\Pi^T$. 
EM duality then gives the poles and zeros for $C_{tt}(w,q)\propto 1/\Pi^T$, as the zeros and poles of
$\Pi^T$, respectively. The low lying poles and
zeros can be found using a variety of different methods. Most crudely,
one can use the direct numerical solution to the EoM, but this turns out to be unstable
as one probes frequencies deeper in the LHP. Alternatively, in Appendix~\ref{sec:heun} we provide a solution for the
correlators in terms of the local Heun function and its derivative. As no closed-form of its series
representation is generally known, and many of its properties are still under study, we were not able
to use it to obtain the QNM spectrum. Notwithstanding, its series representation, for which we give the recursion relation,
can be useful to obtain a solution when the direct solution of the ODE fails. 
We mainly resort to a method that focuses on the QNM spectrum specifically. 
%Let us first discuss the procedure used to find the poles of $\Pi^T$.
It consists in expanding $A_y$ in a Taylor series in the 
radial coordinate $u$, being careful to impose the correct asymptotics at the horizon, and
especially at the UV boundary, $u=0$, where we require $A_y$ to vanish. We then transform the 
EoM into a homogeneous matrix equation and ask for values of $(w,q)$ at which it has a
solution (by finding the points at which one of the eigenvalues vanishes for e.g.). These are the locations of the QNMs.
For further details, see Ref.~\onlinecite{ws}. 
\begin{figure}
\centering%
\includegraphics[scale=.6]{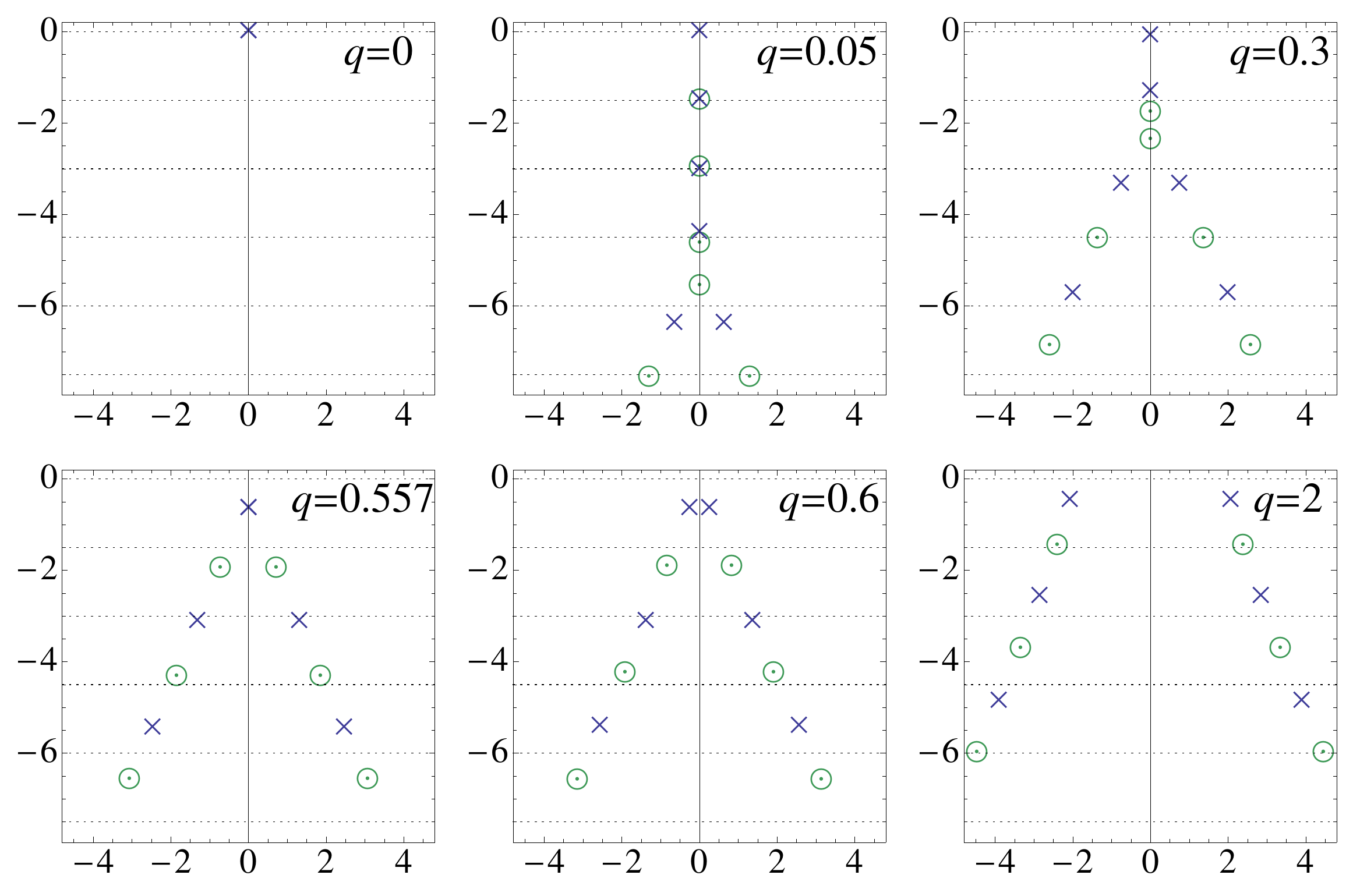}%.48   
\caption{\label{fig:qnm-vs-q} Poles (crosses) and zeros (circles) in the lower-half complex frequency plane
of $C_{tt}=-\chi^2 q^2/\Pi^T(w,q)$,
the R-charge density correlation function of the $\mc N=8$ supersymmetric Yang-Mills CFT. 
The positions of the poles and zeros are interchanged for $C_{yy}=\Pi^T$.
} 
\end{figure} 

The poles and zeros of $C_{tt}(w,q)$ are shown in the LHP of frequency in \rfig{qnm-vs-q}
for six different momenta. At zero momentum, from the exact solution we know that $\Pi^T(w,0)=-i\chi_0w$,
so that $C_{tt}$ has a single pole at the origin. This is the so-called \emph{hydrodynamic pole}. As momentum
is turned on, it disperses quadratically, $w= -iq^2$, as we illustrate in \rfig{crossover}.
This is the hallmark of diffusive behavior.\footnote{The
diffusion equation is $\pd_t\psi=\nabla_x^2\psi$.} The coefficient of $-iq^2$ is precisely $1$.
Indeed, in terms of the unscaled variables, the dispersion relation reads $\w=-iD_0k^2$, where
$D_0=3/4\pi T$ is the charge diffusion constant and $(w,q)=(\w,k)\times D_0$. It is the most important
QNM of $C_{tt}$ as long as it remains bound to the imaginary axis, i.e.\ when $q\leq q_c=0.5573187$,
because in that case it is the QNM with the smallest absolute value of the imaginary part, 
hence it gives the correlation
function its dominant, i.e.\ largest, decay time-scale: $\sim 1/T|\Im w_{\rm QNM}|$.

\begin{figure}
\centering%
\includegraphics[scale=.38]{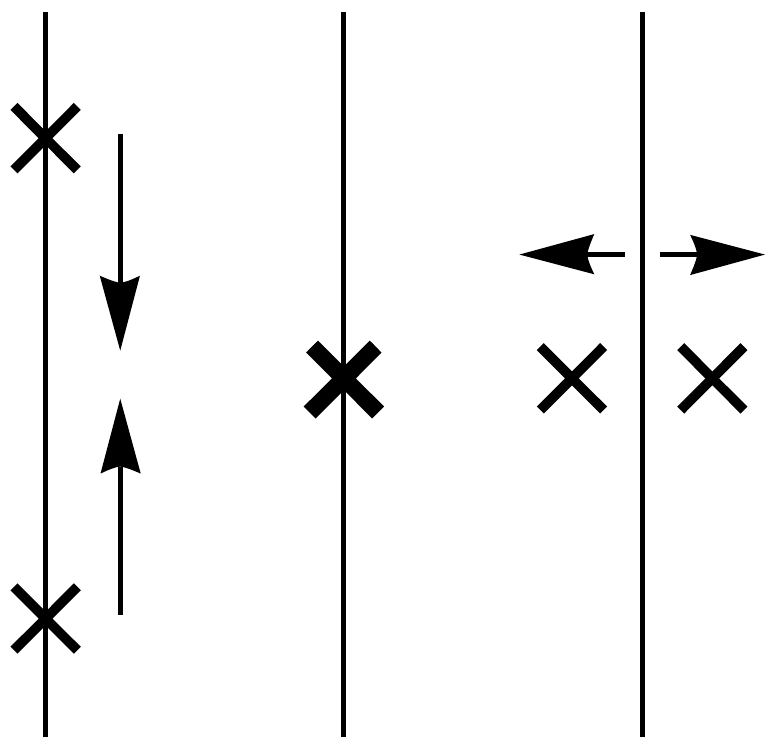}%.48   
\caption{\label{fig:zip} General mechanism according to which two poles detach from
the imaginary axis. A double pole exists at the intermediate step. The same mechanism
applies to zeros. 
} 
\end{figure}

Another important phenomenon occurs at $q\sim 0^+$: pairs of simple poles and zeros nucleate
on the imaginary $w$-axis in the immediate vicinity of the frequencies 
\begin{align}
w_n^{\rm zip}=-i3n/2 \,,
\end{align}
where $n$ is a positive integer. For the unscaled frequency, $\w=4\pi T/3$, 
these correspond to the negative bosonic Matsubara frequencies, $\w_n^{\rm zip}=-2\pi nT$. 
As discussed in Appendix~\ref{sec:heun}, they correspond to known singular points of the local Heun function,
the special function that solves the EoM for $A_y$.
From \rfig{qnm-vs-q}, we see that, with the exception of the special hydrodynamic pole, 
the nucleation ensures that each pole of $C_{tt}$ comes
with a corresponding partner zero, which is a pole for the transverse response function $C_{yy}$.
As the momentum is increased from zero, all these QNMs ``unzip'' from the imaginary axis
as is shown in \rfig{qnm-vs-q}. The unzipping procedure follows the simple rule:
2 poles/zeros join on the imaginary axis to make a double pole/zero, and can
subsequently detach. This elementary mechanism, which is illustrated in \rfig{zip}, is strongly
constrained by time-reversal symmetry, which requires the poles and zeros to be distributed
symmetrically about the imaginary axis. 
A related important property of the zeros and poles is their ordering (according to their norm): 
2 consecutive poles are followed by 2 consecutive zeros, seemingly \emph{ad infinitum}. We note
that a double pole/zero is a superposition of 2 simple poles/zeros and as such respects
the ordering property. In Appendix~\ref{ap:qnm}, we substantiate the claim according to which
double poles or zeros, and not higher order ones, occur in the unzipping process.
We have observed\cite{ws} the same ``unzipping'' phenomenon in the study of the QNMs of the conductivity in 
the four-derivative holographic model which
we discuss below. In that case, the so-called $\g$-coupling plays a role analogous to momentum here.
We indeed expect such a phenomenon to be quite general
as the poles and zeros must be created/destroyed in pairs. 

We observe that it is not only the hydrodynamic QNM that disperses quadratically with momentum
but all the zeros and poles do as well, for sufficiently small $q$. Numerical evidence for this is presented in
\rfig{all-quad} in Appendix~\ref{ap:qnm}.
Here the smallness condition requires the pole-zero pair emanating from a given frequency $w_n^{\rm zip}$ to be
bound to the imaginary axis. The ``dispersion relation'' for a pair associated with $w_{n>0}^{\rm zip}$
is:
\begin{align}
  w = w_n^{\rm zip} \pm i\al_n q^2\,, \quad q\ll 1\,, 
\end{align}
where $\al_n>0$, and $\al_0=1$ as stated above. We have found that  
the dispersion coefficients increase exponentially with $n$ to
good accuracy for $n\geq 2$: $\al_n\approx a_1 e^{a_2 n}$, where $(a_1,a_2)\approx(0.27,1.7)$. 
This leads to the ``unzipping'' process to occur exponentially fast as momentum is increased from zero
so that the QNMs acquire a finite real part very rapidly, as can be observed in \rfig{qnm-vs-q}. 

\subsubsection{Hydrodynamic-to-relativistic crossover} 
\begin{figure}
\centering%
\includegraphics[scale=.63]{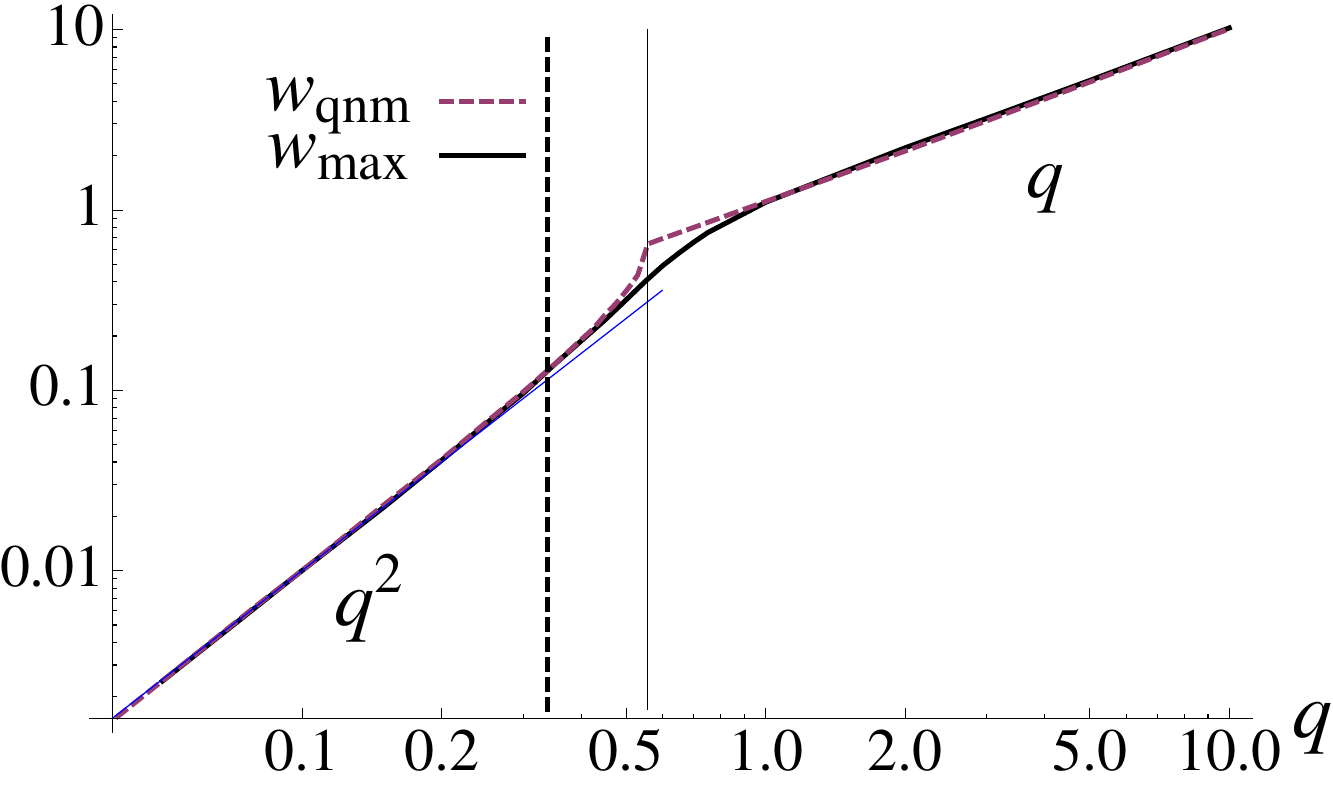}%.48   
\caption{\label{fig:crossover} Hydrodynamic-to-relativistic crossover.
Momentum dispersion of the position of the peak of $C_{tt}(w,q)$, $w_{\rm max}(q)$,
and the norm of the diffusive QNM, $w_{\rm qnm}(q)$. A quadratic hydrodynamic scaling is seen
at small $q$ (as a guide, the thin blue line is $q^2$), while a linear relativistic scaling emerges at large $q$. The vertical lines at $q=0.339$ and $q=0.557$ signal momenta at which pairs of QNMs
detach from the imaginary axis. The second one is where the diffusive QNM detaches.
} 
\end{figure} 
The motion of the QNMs can be used to identify the crossover from hydrodynamic-like behavior
to a relativistic one. (Note that we use the designation ``relativistic'' instead of
``collision-less'', which is sometimes employed, because the latter could suggest the presence of well-defined quasiparticles
interacting with each other whereas such a picture does not hold for the strongly correlated theories we
describe.) 
Let us examine the charge density correlator $C_{tt}$.
For fixed momentum $q$, $-\Im C_{tt}/q^2$ has a peak at a frequency $w_{\rm max}(q)$, as is illustrated in \rfig{ImCtt}.
It was previously observed\cite{m2cft} that $w_{\rm max}(q)\sim q^2$ at small momentum $q\ll 1$, while it 
scales linearly for $q\gg 1$, see \rfig{crossover}. This signals a crossover from a hydrodynamic
behavior at small $q$ to a relativistic one at large $q$. We observe that
this crossover corresponds to sharp transitions in the QNM configuration. 
Namely, from \rfig{crossover} we find that when the momentum reaches $q\approx 0.33$, the location of the peak starts
to noticeably deviate in excess from $q^2$. In the LHP, this actually corresponds to the point 
where a pair of zero QNMs detaches from the imaginary axis, which occurs at $q=0.339328$.
Next, the scaling for $w_{\rm max}(q)$ has an inflection point near $q\approx 0.6$ after which
it rapidly becomes linear. Now,
this can be put in correspondence with the momentum at which the diffusive QNM 
detaches itself from the imaginary axis, which occurs at $q= 0.5573187$.
We further find that the value of $w_{\rm max}(q)$ agrees very well with the norm of the
lowest lying QNM away from the transition region, $q\approx 0.557$, as \rfig{crossover} testifies.

Again, just as the diffusive scaling $q^2$ held true for all the QNMs at sufficiently small momentum,
so does the linear scaling for $q> 1$. Indeed, the absolute value of the real part of the QNMs
grows linearly with increasing momentum, as expected for the low temperature excitations
of a CFT. Moreover, the imaginary part of all the modes approaches zero; we illustrate this
for the one closest to the real axis in \rfig{inv-life}. 
In Appendix~\ref{ap:qnm}, we discuss the rate at which this happens as a function of $q$, which
for the mode closest to the real axis seems to occur slower than $1/q^{1/4}$. Nevertheless,
the lifetime (the inverse of the imaginary part), becomes much less than the excitation energy (real part)
at large momenta, and we can thus interpret the QNMs as quasiparticle-like. 
We also remark that the distance or spacing between theses QNMs decreases as $q$ grows. It is thus 
suggestive that the QNMs evolve towards the formation of branch cuts, which exist at $T=0$
due to the form $C_{tt}\propto 1/\sqrt{-w^2+q^2}$. It should be noted that the QNM spectrum has poles closest to the real axis,
and these would precisely become the branch poles at zero temperature.

\section{General response: beyond Einstein-Maxwell}
\label{sec:gamma}

\begin{figure}
\centering%
\includegraphics[scale=.75]{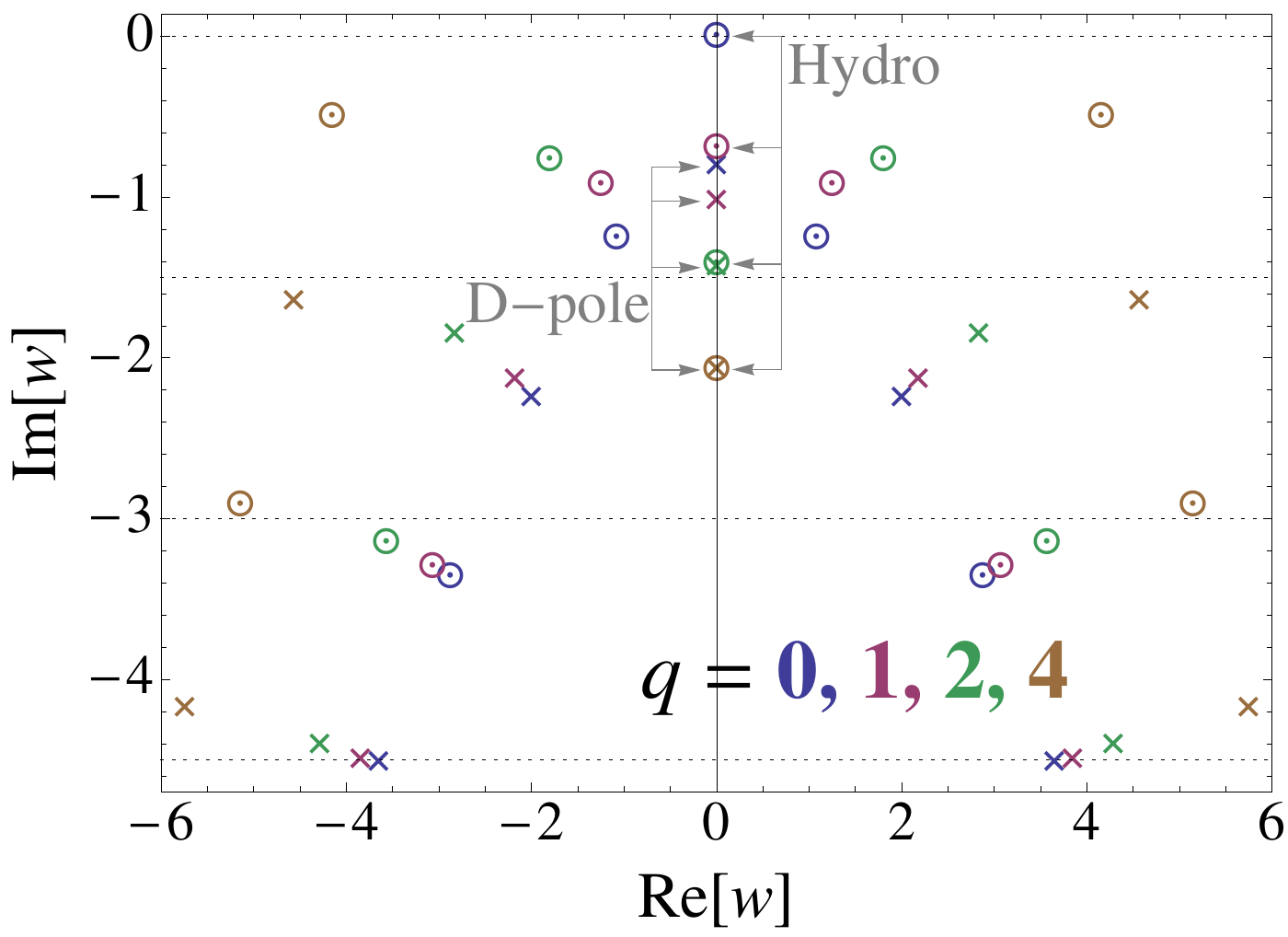}
\caption{\label{fig:PiT-qnm-g083}
QNM dispersion of $C_{yy}=\Pi^T(w,q)$ for $\g=1/12$: the crosses/circles represent
poles/zeros. The $q=0$ case gives the QNMs of the conductivity $\s(w)$, in addition to the hydrodynamic zero at the
origin. The D-pole is the special purely damped QNM giving rise, at $q=0$, to the DS peak in the small-frequency optical conductivity. 
}   
\end{figure}  
We now turn to holographic models which do not possess EM duality, and as such  
have independent transverse and longitudinal responses, as is expected for generic
CFTs. The role of four-derivative terms in the gravitational action on the charge response was
previously considered\cite{ritz,myers11} in an effective field theory spirit. 
The corresponding terms were considered at finite, non-perturbative coupling,
in which case the resulting action cannot be interpreted 
as describing a precise deformation of the original $\mc N=8$ Yang-Mills gauge theory. 
One expects that at finite t'Hooft coupling, $\lambda$,
corrections to the $\lambda=\infty$ limit will include higher-derivative corrections in the gravity side, 
such as the one we consider.
However, considering finite 4-derivative couplings in that context would also entail the need to look at higher order
terms as well, potentially an infinite tower of them.
We adopt an effective approach, considering a truncated action with phenomenological 
couplings parametrizing all symmetry-allowed terms with a fixed number of derivatives.
Based on symmetry, and comparisons with calculations on the CFT side\cite{ws}, we expect this approach to capture 
some salient properties of the correlators.

It was found that in order to observe non-trivial effects on the correlation function of
a single U(1)-current, the only\footnote{We note that a time-reversal symmetric 
term with a lower number of derivatives can appear in
the case of non-abelian symmetries, ``$F^3$'', and it can affect the charge
response when a background charge density is present.}
four-derivative term that needs to be added to the Einstein-Maxwell theory studied in the
previous section, \req{S_bulk}, while preserving time-reversal is
\begin{align}\label{eq:4-deri}
  -\int d^4x\sqrt{-g}\,\g \frac{L^2}{g_4^2}C^{abcd}F_{ab}F_{cd}\,, 
\end{align}
where $C$ is the conformal Weyl tensor, the traceless part of the Riemann tensor, and $\g$ a dimensionless coupling
that was argued to be bounded to $|\g|\leq 1/12$ (we refer to Ref.~\onlinecite{myers11}
and references therein for
a detailed discussion/derivation of the bounds). It was shown that such a term can lead to non-trivial
and generic behavior of the conductivity in contrast to the $\g=0$ frequency-independence. 
In particular, when $\g>0$ $(<0)$ the frequency dependent
charge conductivity is particle-like (vortex-like), with the real part showing a peak (valley) near zero
frequency.  As the Weyl tensor vanishes in pure AdS, this term disappears in the $T\ra 0$ limit,
and will thus not affect the correlators in the relativistic limit of $\w,k\gg T$.
For instance, the $T=0$ conductivity $\s_\infty=1/g_4^2$ is independent of $\g$, taking the same
value as in the $\g=0$ theory.

At finite $\g$, the EM self-duality is absent and an EM duality-transformation leads to a non-trivial action on
the theory, which should manifest itself as S-duality in the boundary CFT. The action on the gravitational
theory is as follows\cite{myers11}: one introduces a term 
$\tfrac{1}{2}\varepsilon^{abcd}\hat A_a \pd_bF_{cd}$ to the Lagrangian, and a functional integral over
$\hat A_a$. $\varepsilon^{abcd}$ is the fully antisymmetric tensor. Such
an addition leaves the partition function invariant since the gauge field 
satisfies the Bianchi identity $\varepsilon^{abcd}\pd_b F_{cd}=0$. Integrating out the
original gauge field yields a new action for $\hat A_a$, which has field
strength $\hat F$:
\begin{align}\label{eq:Sdual}
  \hat S_{\rm bulk}=\int d^4x \sqrt{-g}\left( -\frac{1}{8\hat g_4^2}\hat F_{ab}\hat X^{abcd}\hat F_{cd} \right)\,,
\end{align}
where the tensor $\hat X_{ab}^{\;\;\;cd}=-\tfrac{1}{4}\varepsilon_{ab}^{\;\;\;ef}(X\inv)_{ef}^{\;\;\;gh}
\varepsilon_{gh}^{\;\;\;cd}$  
characterizes the bulk action for the dual gauge field just as $X$ does for $A_a$: $X^{abcd}$ 
simply gives the original action with the Maxwell term and the additional four-derivative $\g$-term, \req{4-deri}.  
The conductivity obtained from the EM-dual gravitational description gives rise to
a conductivity that is the inverse of the original one, $\hat\s=1/\s$, where $\s$ is the
full complex conductivity. This is analogous to what happens under
particle-vortex duality in the O(2) model for instance. It was shown\cite{myers11} that one can relate the current correlators
of the original and S-dual theories as follows:
\begin{align}
  \Pi^L(w,q)\hat\Pi^T(w,q) &=\chi_0\hat\chi_0(-w^2+q^2)\,,\\
  \hat\Pi^L(w,q)\Pi^T(w,q) &=\chi_0\hat\chi_0(-w^2+q^2)\,, \label{eq:hatL-T}
\end{align}
where the hats denote the S-dual correlators, and $\chi_0\hat\chi_0=(4\pi T/3g_4^2)(4\pi T/3\hat g_4^2)=(4\pi T/3)^2$
since $\hat g_4=1/g_4$. The above two relations relate the transverse response of the original theory to the
longitudinal response of the S-dual one and vice-versa.
In particular, the poles and zeros of $\Pi^{L,T}$ map to the zeros and poles of $\hat\Pi^{T,L}$, 
respectively. Further, all
the information is contained in the two transverse correlators $\Pi^T$ and $\hat\Pi^{T}$, which we found
the easiest to compute given the similarity between the EoMs of $A_y$ and its EM dual, $\hat A_y$.
The modified Maxwell equations that we need to solve are $\nabla_b(X^{abcd}F_{cd}) =\nabla_b(F^{ab}-4\g L^2C^{abcd}F_{cd})=0$,
and $\nabla_b(\hat X^{abcd}\hat F_{cd}) =0$,
and they lead to the following EoMs for the transverse gauge field and its EM dual\cite{myers11}:
\begin{align}\label{eq:eomAy-g}
  A_y''+\left(\frac{f'}{f}+\frac{g'}{g}\right)A_y' + \frac{w^2-q^2f(1-8\g u^3)/g}{f^2}A_y &=0\,, \\
  \hat A_y''+\left(\frac{f'}{f}+\frac{g'}{g}\right)\hat A_y' + \frac{w^2-q^2fg/(1-8\g u^3)}{f^2}\hat A_y &=0\,, \label{eq:eomAy-g-dual}
\end{align}
where $g(u)=1+4\g u^3$. They both reduce to \req{eomAy} when $\g=0$, in agreement with self-duality.  

\subsubsection{Sign of $\g$ and EM/S-duality}
From the observations made above we can draw a connection between the sign of $\g$ and the action
of EM duality on the bulk action, and the corresponding S-duality on the boundary. It was previously
noted\cite{myers11,ws} that for $|\g|\ll 1$, the action of the EM duality described above is tantamount to changing the sign
of $\g$. (We note that the inversion of the bulk gauge coupling $g_4\ra g_4\inv$ is of little importance to
our discussion.) This agrees with the particle- and vortex-like conductivities at $\g>0$ and $\g<0$, respectively. 
For general $|\g|\leq 1/12$, this correspondence qualitatively holds although the quantitative agreement 
deteriorates with increasing $|\g|$. Notwithstanding, the conductivity will invariably have a purely imaginary
pole for $\g>0$ and a zero when $\g$'s sign is reversed. We designate these as D-QNMs due to their purely damped
nature as well as their formal relation to the standard Drude form for the optical conductivity. 

\subsection{Finite $\g$ QNMs}
Let us first examine the QNMs of $C_{yy}=\Pi^T(w,q)$ in the particle-like theory 
at $\g=1/12$, as shown in \rfig{PiT-qnm-g083}. 
At $q=0$, this is precisely the meromorphic structure of the conductivity $\s(w)=iD_0\Pi^T(w,0)/w$, with the
addition of the hydrodynamic zero at the origin, which is annihilated by the factor of $1/w$ in the
expression for the conductivity.  
An important difference with the $\g=0$ case 
discussed above (\rfig{qnm-vs-q}) is that even at $q=0$, $\Pi^T$ already has a sequence of poles and zeros
lying away from the imaginary axis. Another important difference is the presence of a pole 
directly on the imaginary axis at $w_{\rm D}(q=0)=-i0.821075$, this is the D-pole of the conductivity
discussed previously\cite{will,ws}. Such a pole is absent in the $\g=0$ self-dual theory and alters the QNM
spectrum in an essential way. In particular, it can lead to a different kind of crossover in
the spectrum compared with the $\g=0$ theory. However, before discussing the intermediate crossover regime, $q\sim 1$,
let us examine what happens at small momenta $q\ll 1$.% and $q\gg 1$.

\subsection{Hydrodynamic zero and S-dual diffusion constant}
As the momentum $q$ is increased from 0, both the hydrodynamic zero and
the D-pole will move down the imaginary axis. Already at $q=2$, they are almost on top of each other (\rfig{PiT-qnm-g083}), and 
as momentum is increased they further move down as a tightly bound pair. At small momentum,
the hydrodynamic zero disperses as 
$w=-i0.625 q^2$ while the D-pole as $w=w_{\rm D}(q=0)-i 0.2264q^2$.
It is interesting to note that since the hydrodynamic zero disperses faster than the D-pole, 
they will eventually fuse (at which point they momentarily disappear) and then move through each other. 
This happens when $q\approx 2.1215$, and the crossing or fusing frequency
is found to be precisely the first zipping point $w_1^{\rm zip}=-i3/2$. We now take a closer look into the small momentum
dispersion of the hydrodynamic zero. 

We recall \req{Cyy-hydro}, which says that at small frequencies and momenta, the transverse correlator becomes
\begin{align}\label{eq:Cyy-hydro2}
  C_{yy}(\w,k)= -\s_{\rm dc}(i\w-\hat{ D}k^2)\,,
  % -\frac{\s_{\rm dc}{D_0}(iw-\hat{\mc D}q^2)
\end{align}
where $\hat D$ is the charge diffusion constant of the S-dual theory;
%where $\hat{\mc D}=\hat D/D_0$ is the charge diffusion constant of the S-dual theory, 
%$\hat D$, divided by $D_0=3/4\pi T$, the diffusion constant of the $\g=0$ theory; 
$\s_{\rm dc}=(1+4\g)g_4^{-2}$ is the d.c.\ conductivity\cite{myers11}.
%$\s_0=1+4\g$ is the d.c.\ conductivity.
The appearance of the S-dual diffusion constant and not $D$
can be seen to arise from \req{hatL-T}, which relates the transverse response to 
the inverse longitudinal response of the S-dual theory. As mentioned above, the dispersion
of the zero has yielded $\hat D/D_0\approx 0.625$, which
is close but not equal to $D_0/D(\g=1/12)=0.579$, so that $\hat{D}\neq 1/D$
and the relation between the diffusion constants (and hence the susceptibilities) is not as
simple as that between the d.c.\ conductivities. As matter of fact, $\hat{D}$ is closer
to $D(\g=-1/12)=0.585D_0$.

We can use the extension\cite{kovtun-stretch,brigante} of the membrane paradigm\cite{thorne} adapted to our 
gravitational action to determine the actual diffusion constant of the S-dual theory. 
The general idea is to consider a stretched horizon located at $r_s$, with $r_s>r_0$.
One then combines the stretching-direction 4-vector $n_\mu=(0,0,0,g_{rr}^{1/2}r_s)$ with the field strength
to form a conserved current, $j^\mu=n_\nu F^{\mu\nu}$. The conservation law for the latter can be recast as a diffusion equation, $\pd_tj^0=\hat D\pd_i\pd_i j^0$, where $\hat D$ is the charge diffusion constant we seek.
Adapting the expression\cite{myers11} for the diffusion constant to the dual theory we get 
\begin{align}
  \hat D=D_0\sqrt{-g}\left.\sqrt{\smash[b]{-\hat X^{xtxt}\hat X^{xuxu}}}\right|_{u=1} \int_0^1 \frac{du}{\sqrt{-g}\hat X^{tutu}}\,,
\end{align}
where the tensor $\hat X$ describes the action for the dual gauge field, as introduced in \req{Sdual}.
From Ref.~\onlinecite{myers11} we have $\hat X_{tx}^{\;\;\;tx}=\hat X_{xu}^{\;\;\;xu}=1/(1+4\g u^3)$ and 
$\hat X_{tu}^{\;\;\;tu}=1/(1-8\g u^3)$.
Finally, using our AdS-Schwarzchild metric \req{metric} %$g_{ab}=u^{-2}\diag(-f,1,1,1/f)$,
we obtain
\begin{align}\label{eq:hatD-ans}
 \frac{\hat D}{D_0}=\frac{1-2\g}{1+4\g}\,.
\end{align}
Evaluating this expression at $\g=1/12$, we find $\hat D/D_0=5/8=0.625$, in exact agreement
with our above numerical result for the dispersion relation of the hydrodynamic zero of $\Pi^T$.
We note that the dual diffusion constant takes a simpler form than in the direct theory, where
it reads\cite{myers11}
\begin{align}
  \frac{D}{D_0}=\frac{1+4\g}{12\g^{1/3}}\left[ \pi\sqrt{3} 
-2\sqrt 3 \tan\inv\left(\frac{1+\g^{1/3}}{\sqrt 3 \g^{1/3}} \right) +\ln\left( \frac{1-8\g}{(1-2\g^{1/3})^3}\right) \right]\,.
\end{align}
\begin{figure}
\centering%
\includegraphics[scale=.47]{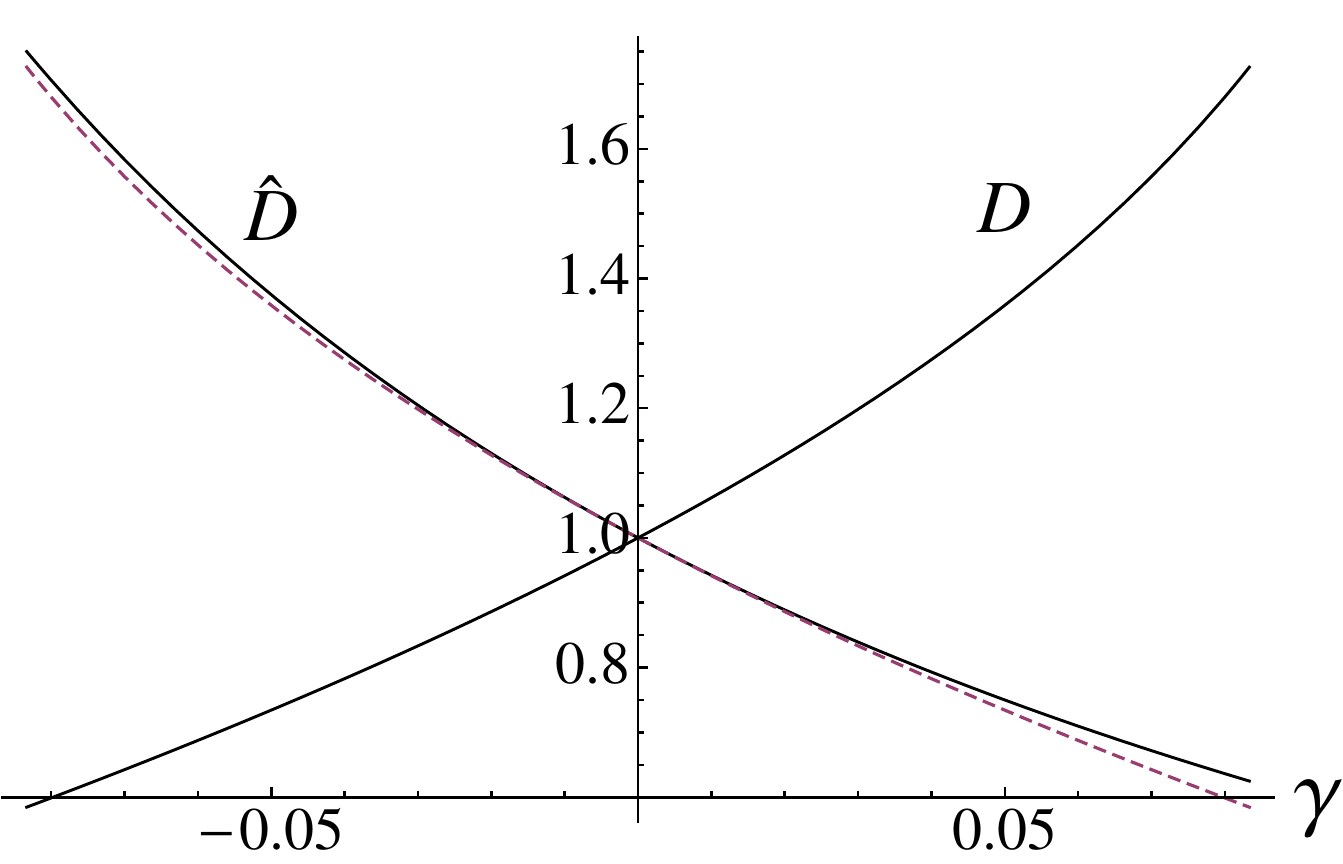}  
\caption{\label{fig:D} Diffusion constant, $D$, and its S-dual, $\hat D$, in units of $D_0$
for $|\g|<1/12$.
The dashed line gives $D(-\g)$, which agrees well with $\hat D$ at small $|\g|$ in accordance
with the relation between the sign of $\g$ and EM/S-duality. Although not shown, $1/D$ is very
close to $D(-\g)$ and hence to $\hat D$.
} 
\end{figure}
A plot of the diffusion constants $D$ and $\hat D$ can be found in \rfig{D}. 

One can also apply an Ohm's law to the stretched horizon\cite{ritz} to recover the S-dual d.c.\
conductivity: 
\begin{align}
  \hat\s_{\rm dc} &=\frac{1}{\hat g_4^2}\left.\sqrt{-g}\sqrt{\smash[b]{-\hat X^{xtxt}\hat X^{xuxu}}}\right|_{u=1} \\
  &= \frac{g_4^2}{1+4\g}\,. 
\end{align}
This agrees with the action of S-duality on 
the conductivity: $\hat\s=1/\s$, valid at all frequencies.
Finally, using the Einstein relation for the S-dual theory, $\hat D\hat\chi=\hat\s_0$, we get a
simple expression for the S-dual charge susceptibility: 
\begin{align}
  \hat\chi=\frac{4\pi Tg_4^2}{3} \frac{1}{1-2\g}
\end{align}

\begin{figure}
\centering 
\subfigure[]{\label{fig:gam-cross} \includegraphics[scale=.5]{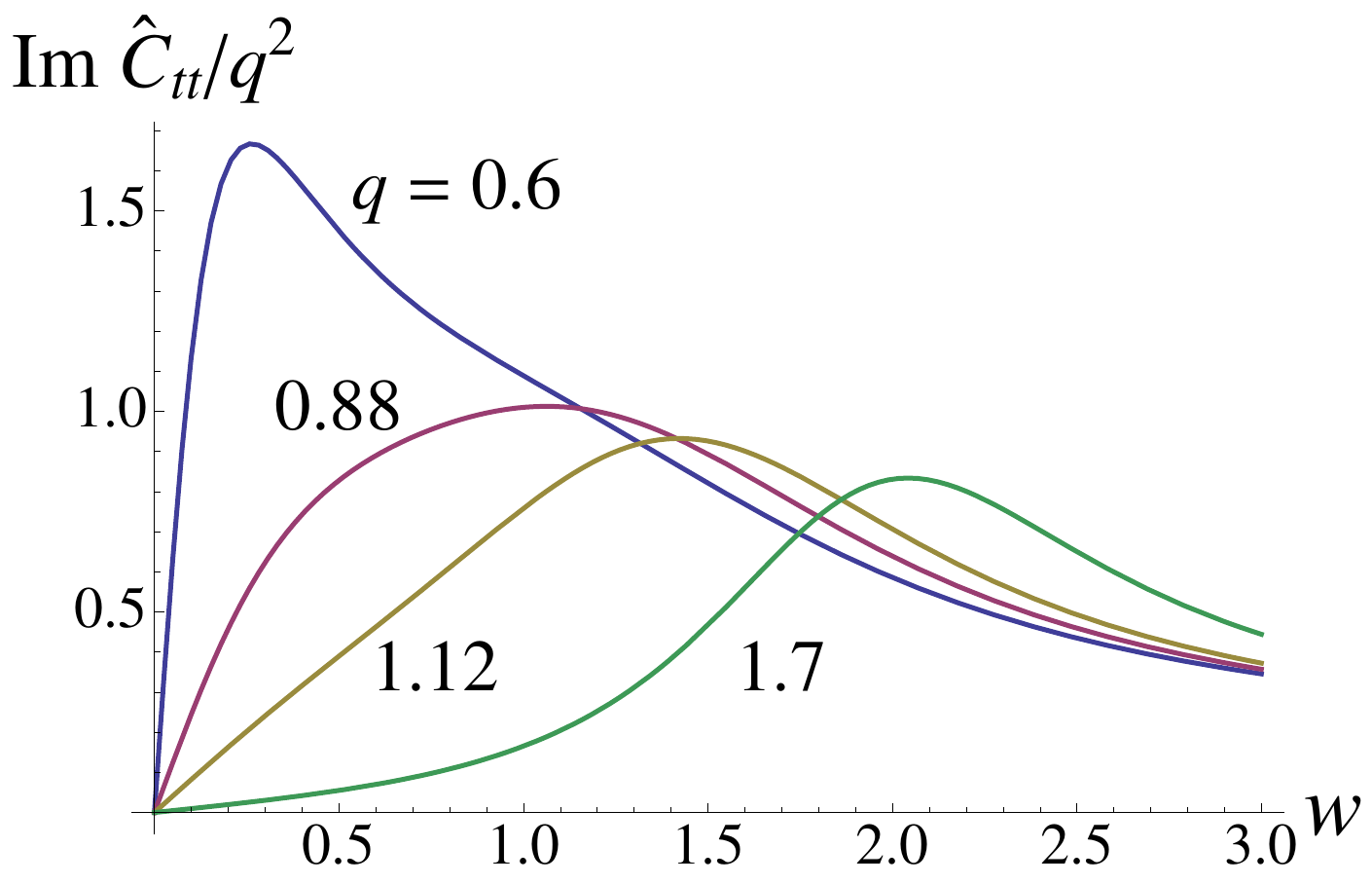}}
\subfigure[]{\label{fig:gam-wmax} \includegraphics[scale=.57]{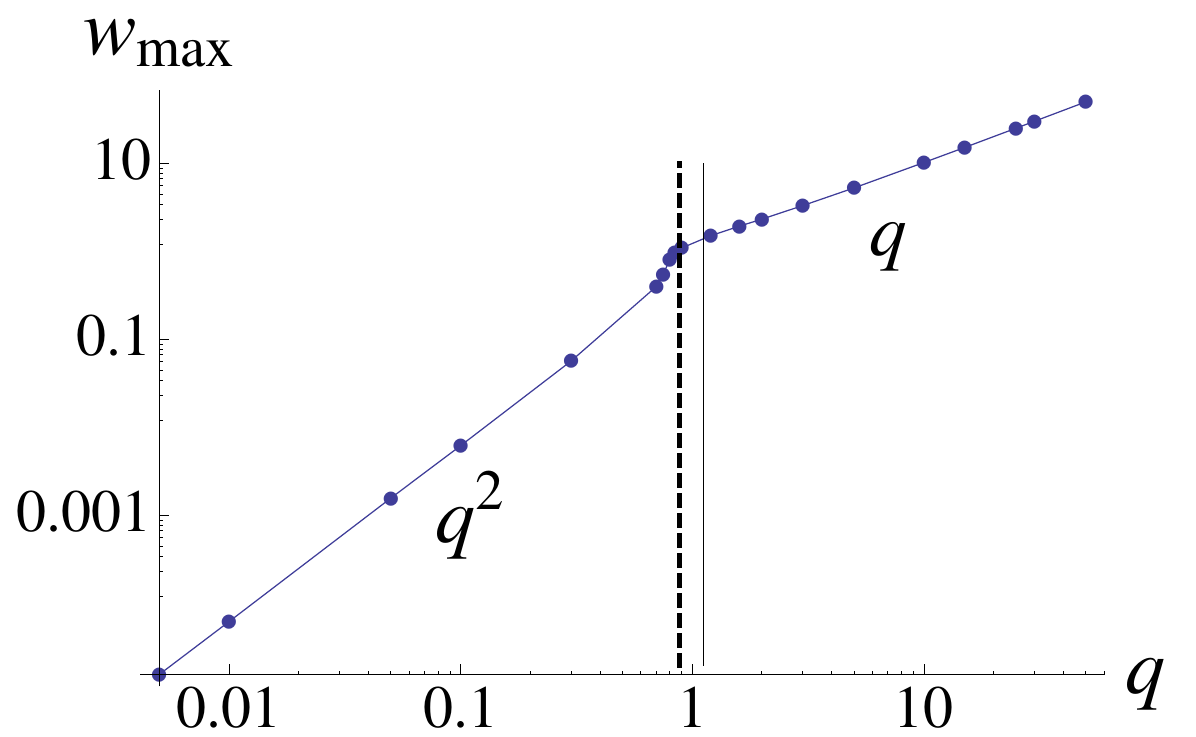}}\\
\subfigure[]{\label{fig:gam-qnm-cross}\includegraphics[scale=.3]{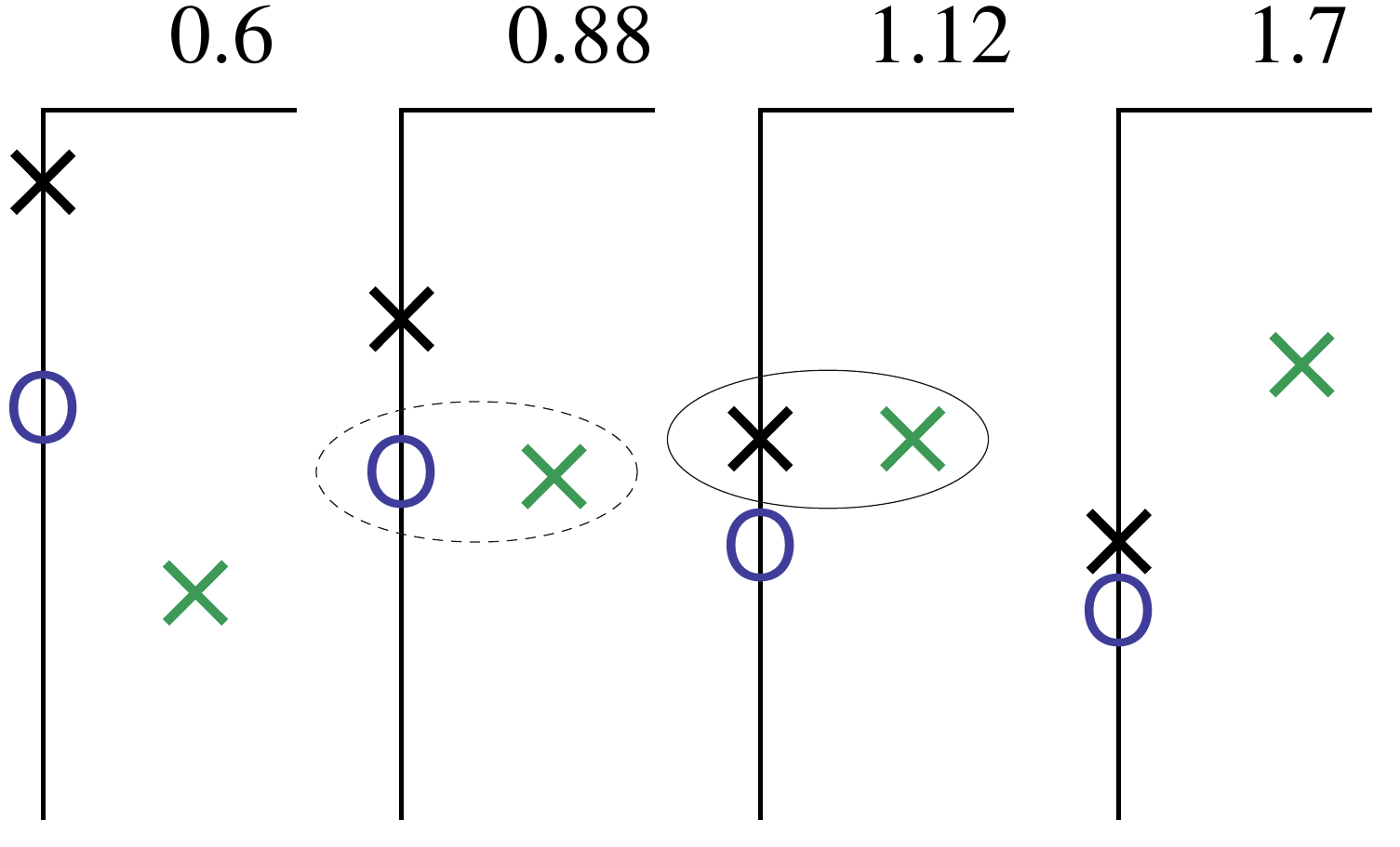}}\hspace{70pt} %\null\hfill
\subfigure[]{\label{fig:cross-2-cases}\includegraphics[scale=.4]{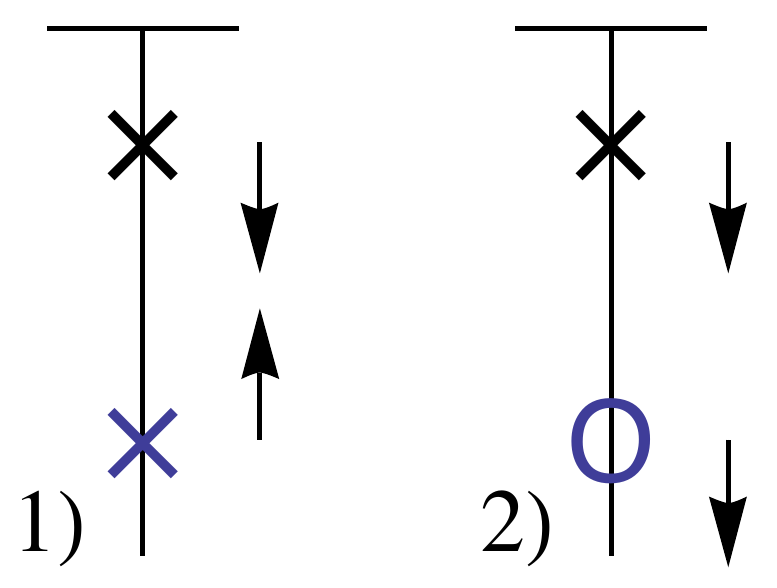}} 
\caption{\label{fig:g-cross} Hydrodynamic-to-relativistic crossovers at finite $\g$. 
a) $\Im \hat C_{tt}(w,q)$ at $\g=1/12$, in units of $-\hat\chi_0$, for 4 different momenta.
b) Dispersion of the peak of $\Im \hat C_{tt}$. The vertical dashed and solid lines correspond to $q_{c1}=0.88$
and $q_{c2}=1.12$, respectively. 
c) The momentum evolution of the 3 QNMs nearest the real axis in the complex $w$-plane ($\Re w\geq 0$ \& $\Im w <0$). 
Poles are represented by crosses while zeros by circles. d) Two cases for the motion of the hydrodynamic pole
and the D-QNM. Case 2 applies to the correlator under study in panels a, b \& c. 
Case 1 applies to $C_{tt}(\g>0)$ for example. 
} 
\end{figure}

\subsection{Hydrodynamic-to-relativistic crossover}
If we examine the behavior of the QNMs away from the real axis, we find that the absolute value of their real part
grows with $q$, where for sufficiently large $q$, the increase is linear with momentum,
just as we found for the $\g=0$ theory. At the same time, the
imaginary part of these propagating QNMs tends to zero. We thus recover relativistic quasiparticle-like QNMs,
and the spectrum evolves towards the asymptotic formation of a pair
of branch cuts emanating from the points $w=\pm q$, characteristic of the $T=0$ form $\sqrt{-w^2+q^2}$.
These phenomena can be observed in \rfig{PiT-qnm-g083}.
At large momenta, $\g$ becomes less important and it is not surprising to
recover behavior similar to the $\g=0$ case. We now explain how the diffusive behavior discussed above
crosses-over to this relativistic regime.

One of the main results of this section is that the presence of a  purely damped \emph{zero}, a D-QNM,
at finite $\g$ can lead to a transition in the QNM spectrum that is distinct from the $\g=0$
case presented above. We make our point using the longitudinal correlator of the S-dual theory: 
$\hat C_{tt}(w,q)\propto 1/\Pi^T(w,q)$ at $\g=1/12$. We can thus make a connection with
the QNM spectrum of $\Pi^T$ given in \rfig{PiT-qnm-g083}. We see that the D-pole of $\Pi^T$ becomes
a D-zero of $\hat C_{tt}$.
We plot the corresponding spectral function for four different 
momenta in \rfig{gam-cross}. The location of the peak scales quadratically
with momentum for $q\lesssim 1$, while the scaling becomes linear in the opposite limit. This
can be seen more clearly in \rfig{gam-wmax}. We also show a sketch of the
evolution of the key QNMs of $\hat C_{tt}\propto 1/\Pi^T$ in \rfig{gam-qnm-cross}, focusing only
on the the QNMs nearest the real axis as they dominate the small frequency response. 
Contrary to the $\g=0$ case, unzipping cannot occur because a pole and zero cannot conspire to detach
from the imaginary axis. In principle, they could annihilate but it turns out this does not occur.
Instead, we see that the linear or relativistic 
behavior begins at $q_{c1}\approx 0.88$ where the purely imaginary zero (the D-pole of $\Pi^T$) 
acquires the same imaginary part as the off-axis pole, see the dashed oval in \rfig{gam-qnm-cross}. 
A secondary crossover
occurs at $q_{c2}\approx 1.12$, where the hydrodynamic pole loses its role of dominance (smallest norm of 
the imaginary part) to the off-axis pole. 
The latter keeps approaching the real axis while its
real part scales linearly with momentum while the purely damped QNMs propagate towards $-i\infty$.
We thus roughly see the general principle at play: the hydrodynamic-to-relativistic crossover occurs when
the off-axis QNM acquires a greater ``lifetime'', $1/|\Im w|$, than the hydrodynamic diffusive mode. 
The presence of the D-mode leads to an intermediate regime, $q_{1c}<q<q_{2c}$, where the 
hydrodynamic QNM dominates yet
a linear scaling of $w_{\rm max}(q)$ can be seen. In this transitory regime, the response shows a broader peak 
signaling the competition of two poles as can be seen in \rfig{gam-cross}. We note that
it is only for momenta in excess of $q_{2c}$ that an inflection point for $\Im C_{tt}$ appears at small frequencies,
allowing for the strong suppression of spectral weight at $w<q$ as momentum increases. 

The example above does not cover all possibilities at finite $\g$, as there are in fact two cases 
depending on whether the
D-QNM is a pole or zero, as is illustrated in \rfig{cross-2-cases}. {\bf Case 2} was the subject
of the preceding paragraph since the D-QNM of $\hat C_{tt}(\g>0)$ is a zero. Generally, case 2 applies to
$C_{tt}(\g<0)$ and $\hat C_{tt}(\g>0)$. To see this it suffices to remember that $\s(w)\propto iwC_{tt}(w,0)/q^2$,
so that D-QNM of the charge correlator arises from the one of the conductivity, and that S-duality
changes a D-pole into a D-zero and vice-versa. {\bf Case 1}, on the other hand, agrees with the $\g=0$ 
theory, with the difference that in the latter situation no D-pole exists but a pole
nonetheless appears at $q>0$ and plays the same role as the D-one in the crossover.
More generally, case 1 applies to $C_{tt}(\g>0)$ and $\hat C_{tt}(\g<0)$.
An important difference outlined in the above discussion is that
in case 1 the poles actually detach from the imaginary axis at some momentum, whereas in case 2 they do not.
An additional and related difference is that the two QNMs move toward each other in case 1, whereas they move
in the same direction in case 2. This will actually cause the crossover to occur earlier, \emph{viz.}\ at a smaller
momentum, in case 1 versus 2 for a fixed value of $\g$. For instance, the two poles of $C_{tt}(\g=1/12)$ 
collide and detach when $q=0.3280$, which corresponds to the point at which the quadratic
scaling of the peak, $w_{\rm max}\sim q^2$, starts crossing-over to a linear one. Note that this is almost
three times less than the value of the critical momentum of $\hat C_{tt}(\g=1/12)$ discussed above
(case 1), 
where we it was found that $q_c\approx 0.9$. 
The same conclusion can be drawn for $\hat C_{tt}(\g=-1/12)$, 
where the unzipping of the hydrodynamic and D-poles occurs at $q=0.3045$. 

\section{Sum rules and causality}
\label{sec:sr}
We discuss certain integral relations involving the current correlators
for any momentum; these include the conductivity sum rules
discussed previously\cite{sum-rules,ws} as special cases. 
As is generally the case when one deals with retarded correlation functions, the
sum rules rely on Kramers-Kronig relations, such as
\begin{align}\label{eq:kk}
  \Im \psi(w') = \frac{1}{\pi}\mc P\int_{-\infty}^\infty dw\frac{\Re\psi(w)}{w-w'} 
\;\; \xrightarrow{\;w'\ra 0} \;\;
  \Im \psi(0) = \frac{1}{\pi}\mc P\int_{-\infty}^\infty dw\frac{\Re\psi(w)}{w}
\end{align}
where $\psi$ is analytic in the UHP and decays sufficiently fast at infinity. The arrow indicates 
the zero-frequency limit $w'\ra 0$ of this Hilbert transform relation, which will be our main tool below.
One subtlety arises because CFTs 
have an abundance of excitations at all energy scales, namely the appearance of UV singularities.
These can be dealt with using the appropriate simple subtractions\cite{sum-rules,ws}. 

The first relation is  
\begin{align}
  \int_0^\infty dw \frac{\Re[i\Pi^T(w,q)-\chi_0 w]}{w} &= -\frac{\pi}{2}\Pi^T(0,q)\,, \label{eq:sr1} \\
 q\ra 0:\;\;\; \int_0^\infty dw [\Re \s(w)-\s_\infty] &= 0\, \label{eq:sr1-sig}\,.
\end{align}
We have omitted the principal value in the first equation because the integrand is finite at $w=0$ since
$\Pi(0,q)$ is real; further, the extra factor of $1/2$ on the r.h.s.\ appears because the integral is over the
non-negative frequencies only.
The static correlator on the r.h.s.\ of the first equation is not only real but also
positive for $q> 0$. It vanishes
identically at zero momentum, $\Pi^T(0,0)=0$, yielding the conductivity sum
rule, \req{sr1-sig}. 
The convergence of the integral in the l.h.s.\ of \req{sr1} is guaranteed by the fact 
that the function $\Re[i\Pi^T(w,q)-\chi_0w]/w$
decays sufficiently fast as $w\ra\infty$ and $q$ remains finite, namely as $1/w^2$. 
Note that this is the slowest 
integer-power decay compatible with the odd nature of $\Im \Pi^T$. This can be understood by referring
to the zero temperature form $-\Im\sqrt{-w^2+q^2}$, to which the finite temperature correlator tends
in the $w\gg q$ limit. Its asymptotic expansion is $-w +q^2/2w+\mc O(w^{-3})$, in accordance with our above claim
for the decay of the integrand.
We have numerically verified that the coefficient of the subleading term, $q^2/2$, matches the 
behavior of the full solution.
It should be noted that the scaling for $q=0$ is generically different. When $\g=0$ for instance, the
spectral function does not have subleading terms as it scales exactly linearly. For finite $\gamma$,
we have numerically found that the subleading term decays faster than $w^{-2}$, although we cannot
at this time soundly establish the precise power.

We now discuss the basic physics underlying the conductivity sum rule \req{sr1-sig}. As 
we just saw, the reasons underlying the Kramers-Kronig relation, \req{sr1}, are
1) the causal structure of the correlation function (analyticity in the UHP of frequency)
and 2) the sufficiently fast decay of the integrand at large frequencies essentially due to Lorentz invariance. 
It remains to understand why does the r.h.s.\ of \req{sr1}, $\propto \Pi^T(0,q)$, vanish at zero momentum,
leading the conductivity sum rule. One way to see this is to recall that the conductivity $\s(w)=i\Pi(w,0)D_0/w$
of the CFT is finite in the d.c.\ limit due to the particle-hole or charge-conjugation symmetry. As such,
$\Pi^T(w,0)\propto -i w$, i.e.\ it must vanish at zero frequency.  
Now, can we also make an argument that relies purely on the holographic bulk? We argue to the positive:
the sum rule is a manifestation of
\emph{gauge invariance of the bulk gauge field}. In the rest of the paragraph, we explain the argument 
connecting bulk gauge invariance with the
vanishing of the transverse correlator at zero frequency and momentum, $\Pi^T(0,0)=0$.
We first recall the AdS/CFT relation $\Pi^T(w,q)=-\chi_0 A_y'(0;w,q)/A_y(0;w,q)$,
where primes denote $u$-derivatives. So that $\Pi^T(0,0)=0$ is equivalent to $A_y'(0;0,0)=0$, 
where we have used the fact that $A_y$ remains finite in the limit under consideration. This
follows from the existence of a well-behaved hydrodynamic limit for $\Pi^T$. 
Further, $A_y'$ vanishes
in the near-horizon region in the limit of vanishing frequency and momentum as is shown at the
end of the paragraph. The EoM for
$A_y$ reads $A_y''+p_1(u)A_y'+p_2(u)A_y=0$, where crucially $p_2=0$ when $w=q=0$ since
by gauge invariance terms without derivatives must vanish (a mass is not allowed). 
The resulting EoM without $p_2A_y$ then simply propagates the property $A_y'|_{w=q=0}=0$
all the way to the UV boundary at $u=0$, and this proves
that $A_y'(0;0,0)=\Pi^T(0,0)=0$. Gauge invariance of the bulk gauge field was essential in the
argument. We now prove our claim according to which $A_y'$ vanishes in the near-boundary region when
$w,q\ra 0$.
Near the horizon $u=1$ one needs to apply
an in-falling boundary condition to solve for the retarded $\Pi^T$: $A_y(u;w,q)=(1-u)^{-iw/3}F(u)$ as $u\ra 1$, where
$F$ is analytic in the vicinity of the horizon and we have the freedom
to set $F(1)=1$. Taking a derivative of $A_y$ gives:
$A_y'(1+\e;w,0)=-i\tfrac{w}{3}(\epsilon\inv+e_1)(-\epsilon)^{-iw/3}$, with $\epsilon=0^+$ so
that $u=1+\e$ is just outside of the horizon;
$e_1$ is a finite number coming from $F'(1)$. We have used the fact that $F'(1)$ vanishes linearly with $w$ when
$q=0$, which is readily seen to rely on the vanishing of $p_2$ at $w=q=0$ 
as guaranteed by gauge invariance\footnote{In the limit of $u\ra 1$, the EoM becomes:
  \begin{align}
    \tilde p_1 F'+[b(b-1)- b\tilde p_1 +\tilde p_2]F=0
  \end{align}
where $\tilde p_n=\lim_{u\ra 1}(1-u)^n p_n(u)$ is finite for both $n=1$ and 2; $b=-iw/3$. As we saw, gauge
invariance requires $\tilde p_2$ to vanish at $w=q=0$. When combined with $F(1)=1$, we obtain $F'(1)\sim w$
at small frequencies.
}. 
Finally taking the limit $w\ra 0$ leads to the desired property. We note that at finite momentum,
both $F'(1)$ and the term $p_2(u)A_y$ remain finite as $w\ra 0$. 
This will lead to $\Pi^T(0,q)$ being finite for $q>0$, and we now turn 
to study that function in more detail.
\begin{figure}
\centering
\includegraphics[scale=.6]{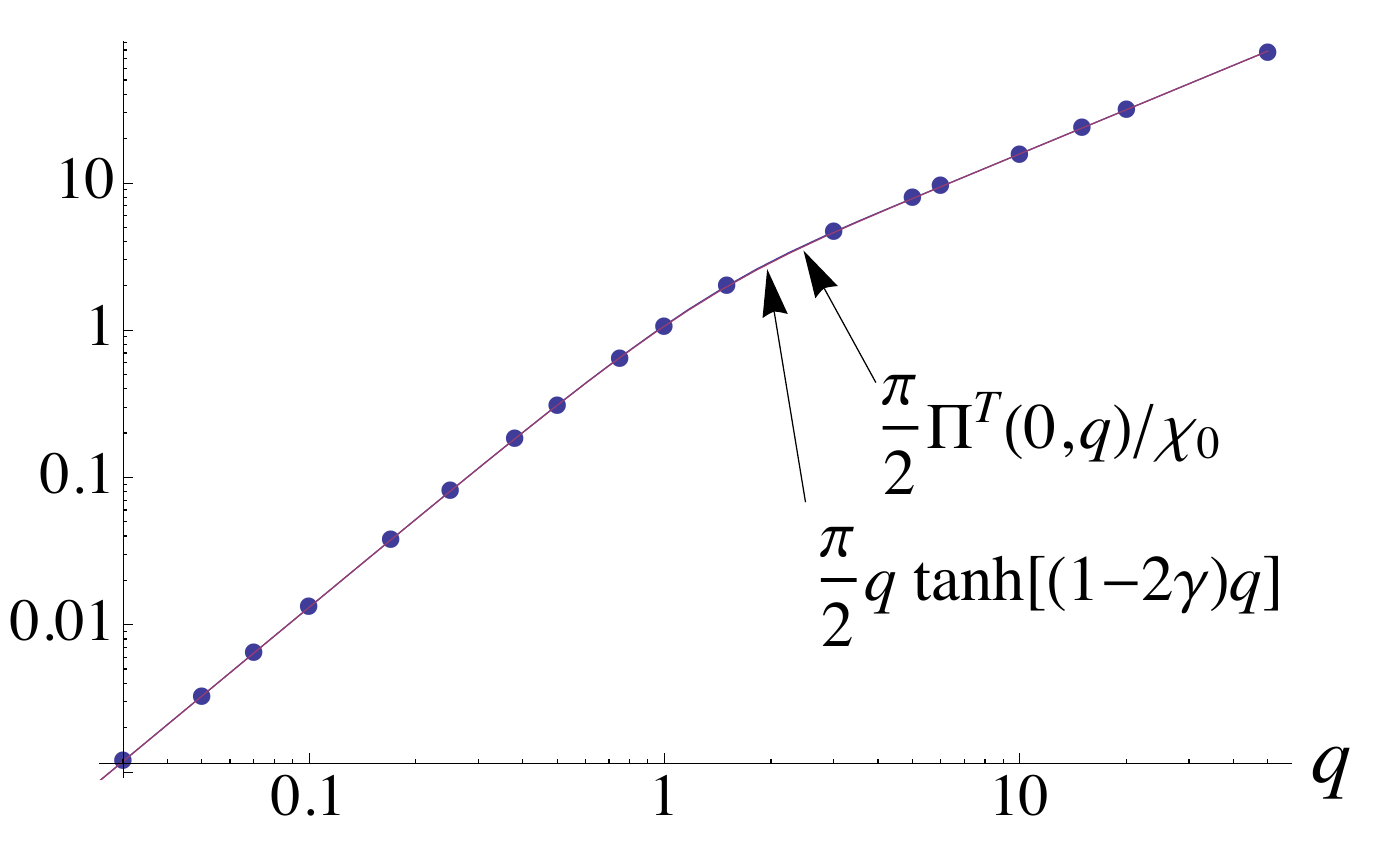} 
\caption{\label{fig:sr} Verification of sum rule \req{sr1} for $\g=1/12$. The dots correspond to the
absolute value of the integral
on the l.h.s.\ of \req{sr1}, while the two lines (which cannot be distinguished here) to
the static correlator on the r.h.s.\ of \req{sr1} and its almost exact analytical form.
}  
\end{figure}

We have found that the static correlator obeys the following form to excellent accuracy
at all momenta:
\begin{align}\label{eq:PiT-static}
  \Pi^T(0,q)=\chi_0 q\tanh\left(\frac{\hat\chi_0}{\hat\chi}q\right)
\end{align} 
where $\hat\chi/\hat\chi_0=1/(1-2\g)$ is the normalized S-dual charge susceptibility. Not only
is the asymptotic behavior exactly captured by that function, the agreement in
the crossover region is also excellent, as is shown in \rfig{sr}.
We note that at small momenta the correlator vanishes quadratically, 
$\Pi^T(0,q)= (\chi_0\hat\chi_0/\hat\chi)q^2=\chi D\hat D k^2$, in agreement with the
hydrodynamic form \req{Cyy-hydro2}, 
which also guarantees the conductivity sum rule since $\Pi^T(0,0)=0$. At momenta greater than the 
temperature, the behavior crosses-over rapidly to a linear scaling independent of the S-dual susceptibility.
We note that precisely the same form arises for the IR fixed point of the vector O($N$) model in the $N\ra \infty$ limit,
as will be discussed in the next section (\rfig{sr-rot} and \req{rot-tanh}).

The second relation is obtained by taking the S-dual of \req{sr1}, with the replacement $\Pi^T(w,q)\ra \hat\Pi^T(w,q)$:
\begin{align}
  \int_0^\infty dw \frac{\Re[i\hat\Pi^T(w,q)-\hat\chi_0 w]}{w} &= -\frac{\pi}{2}\hat\Pi^T(0,q)\,, \label{eq:sr2} \\
 q\ra 0:\;\;\; \int_0^\infty dw [\Re \hat\s(w)-\hat\s_\infty] &= 0\,. \label{eq:sr-sig-dual} % \quad \hat\s(w)=1/\s(w)
\end{align}
In the limit of zero momentum, the S-dual relation \req{sr2} yields the sum rule for the dual conductivity\cite{ws}, 
$\hat\s(w)=1/\s(w)$. The same physical arguments for the sum rule given above apply here as well.
Also, we again find an almost exact expression for the static
correlator appearing in the r.h.s.\ of the integral relation:
$\hat\Pi^T(0,q)\approx\hat\chi_0 q\tanh\left(\frac{\chi_0}{\chi}q\right)$.
This is the S-dual analog of \req{PiT-static}. 

One can also examine another relation that is related to the sum rule for the S-dual conductivity:
\begin{align}\label{eq:sr3}
  \int_0^\infty dw \left\{ \Re\left[\frac{w}{i\Pi^T(w,q)}\right]-\frac{1}{\chi_0} \right\} &= 0 %\\
 % q\ra 0:\;\;\; \int_0^\infty dw \left\{ \Re\left[\frac{1}{\s(w)}\right] -\frac{1}{\s_\infty} \right\} &=0 
\end{align}
Its proof again follows from the Kramers-Kronig relation \req{kk} with $\psi(w)=[w^2/i\Pi^T(w,q)]-w/\chi_0$, which
vanishes at zero frequency for all momenta, $\psi(0)=0$, yielding a momentum
independent r.h.s\ to the above equation, contrary to \req{sr1} and \req{sr2}. The integrand of \req{sr3} again vanishes as
$1/w^2$ as can be easily seen using the asymptotic expansion given above: $\Im \Pi^T\propto -w +q^2/2w+\mc O(w^{-3})$.
The real part decays faster and can be safely neglected here.
Note that \req{sr3} reduces to the sum rule for the S-dual conductivity,
\req{sr-sig-dual} in the $q\ra 0$ limit. Again, the S-dual version also holds and we can use it to
establish a sum rule for the spectral density of the charge correlator:
\begin{align}
  \int_0^\infty dw\left\{-\Im\left[\frac{wD_0^2C_{tt}(w,q)}{q^2} \right]-\frac{1}{\hat\chi_0} \right\}=0
\end{align}

Finally, we note that although we focused on the zero frequency limit of the Kramers-Kronig relations,
they can be used to determine the real part of all the retarded correlators for
all frequencies and momenta using the imaginary part, and vice-versa.

\section{Comparison with the O($N$) model}
\label{sec:on}
In this section, we compare the current correlators of the vector O($N$) NL$\s$M in 2+1D obtained in 
the large-$N$ limit with the holographic results. It is important to note that we do not
claim that such a model has a well-defined classical (super)gravity description as is the
case for the $\mc N=8$ gauge theory discussed above. This is indeed probably not the case.
Rather, we point out that the holographic results capture some essential properties of the current
correlators and provide a useful platform to compare with generic CFTs. 

\begin{figure}
\centering%
\subfigure[~$-\Im C_{yy}$]{\label{fig:rot-Cyy} \includegraphics[scale=.6]{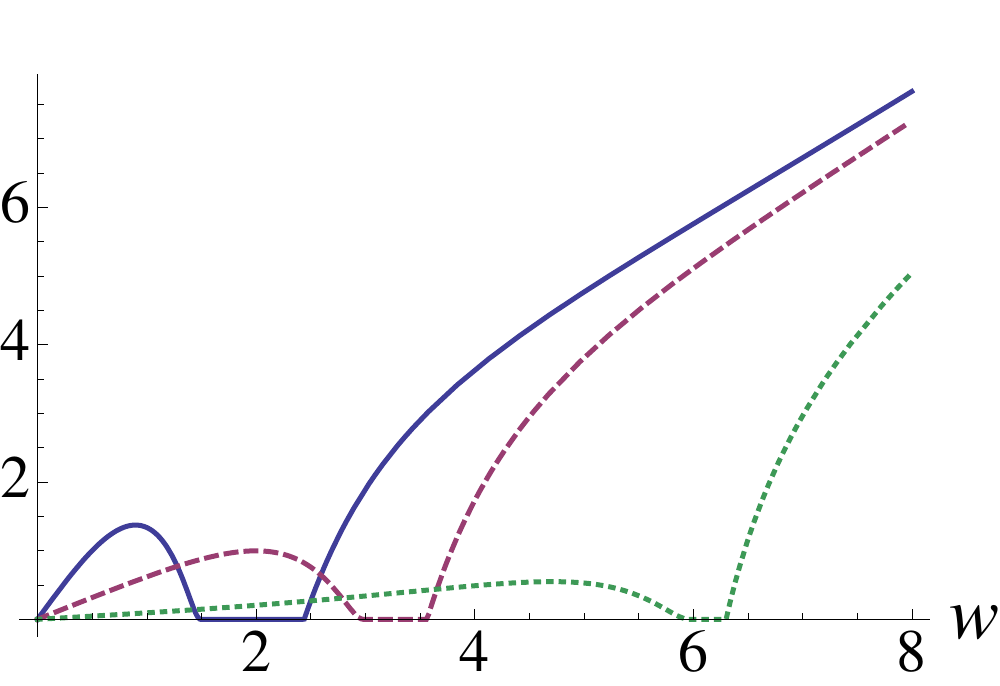}}
\subfigure[~$-\Im C_{tt}/q^2$]{\label{fig:rot-Ctt}\includegraphics[scale=.6]{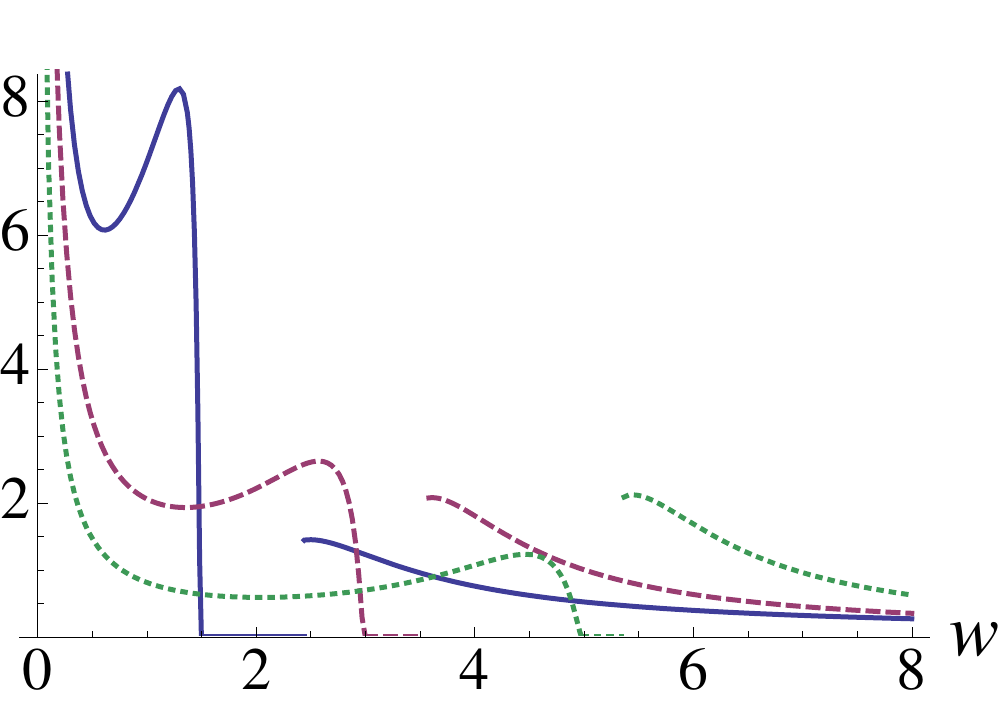}}  
\caption{\label{fig:rot-C} Quantum critical current correlators of the 
vector O($N$) model in the $N\ra \infty$ limit. They have been normalized by $T\s_\infty$, with $\s_\infty=1/16$.
From left to right: $q=1.5$ (solid), 3 (dashed), $6$ (dotted).
} 
\end{figure}

The Lagrangian for the vector O($N$) NL$\s$M is
\begin{align}
  \mc L=\frac{1}{2}\pd_\mu\varphi^a(x)\pd^\mu\varphi^a(x)\,, 
\end{align}
with the constraint $\varphi^a\varphi^a=N/g$, $a=1,\dots,N$ and $g$ is the bare coupling.
At large but finite $N$, the model has a well-known weakly interacting conformal IR fixed point
(for a review and references, see Ref.~\onlinecite{book}),
equivalent to the Wilson-Fisher fixed point accessed by perturbative RG. We study the two-point
functions of the conserved $N\choose 2$ currents near the fixed point, mainly in the $N\ra \infty$ limit. 
Just as in the case of the SO(8) 
R-currents, the correlators are flavor-diagonal and we can focus on a single flavor. 
%\subsection{$N=\infty$}
In the $N\ra\infty$ limit, the fixed point is free, yet 
shows non-trivial dynamics at finite temperature, see \rfig{rot-C} and the discussion below. 
The spatial current correlators can be computed using the one-loop polarization functions in the $N\ra\infty$ limit
(normalized such that $\s_\infty=1/16$): 
\begin{align}
  C_{ij}(\w,\b k)&=\int \frac{d^2\b p}{(2\pi)^2}\frac{p_i p_j}{4\e_p\e_{\b p+\b k}}\left\{ \frac{1+n(\e_p)+n(\e_{\b p+\b k})}{-\e_{\b p+\b k}-\e_p+\w+i0^+} -\frac{1+n(\e_p)+n(\e_{\b p+\b k})}{\e_{\b p+\b k}+\e_p+\w+i0^+}\right.\nn\\
&\qquad \left.+\frac{n(\e_{\b p+\b k})-n(\e_p)}{\e_{\b p+\b k}-
\e_p+\w+i0^+}+\frac{n(\e_p)-n(\e_{\b p+\b k})}{\e_p-\e_{\b p+\b k}+\w+i0^+} -\Re(k\ra 0)
  \right\}\,, \label{eq:rot-C}
\end{align} 
where $\e_k=\sqrt{k^2+m^2}$ and $n(z)=1/(e^{z/T}-1)$ is the Bose-Einstein distribution. 
The ``mass'' parameter is $m=\Theta T$ with $\Theta=2\ln[(1+\sqrt 5)/2]\approx 0.96$;
it corresponds to the temperature-generated mass of the quasiparticles\cite{damle}.
It is simply the inverse correlation length, and must vanish at the QC point. The subtraction in 
\req{rot-C} is necessary to
regulate the UV divergence of the real part. 
In this section,
we use the rescaling $(w,q)=(\w,k)/T$, without the factor of $3/4\pi$ appearing in the holographic
analysis. Just as in the rest of the paper, the momentum
is assumed to lie along the $x$-direction. 

Let us first examine the transverse correlator $C_{yy}$. It can be numerically evaluated 
and the result for the imaginary part is shown in \rfig{rot-Cyy}. 
The behavior of $C_{yy}(w,q)=\Pi^T$ qualitatively resembles 
what has been obtained holographically, the more so in the large-$w,q$ limit where both the vector O($N$) and 
holographic results approach the zero temperature form $\sqrt{-w^2+q^2}$.
However, one important ``pathology'' of the $N=\infty$ limit is the non-meromorphicity of the finite-temperature correlators,
which occurs even at finite momentum. 
Indeed, $\Im C_{yy}$ vanishes identically for $k\leq |\w|\leq \sqrt{4m^2+k^2}$.
The lack of spectral weight in this region is of kinematical origin: the relativistic quasiparticles 
can absorb/emit energy $\w$ and momentum $k$ as long as $\w\leq k$, 
or they can be created in pairs when $\w>\sqrt{(2m)^2+k^2}$, the latter
being the minimal energy for two quasiparticles of mass $m$ carrying total momentum $k$. 
Such constraints become irrelevant as interactions appear at $\mc O(1/N)$ and the spectral weight will thus
become finite throughout when $N<\infty$, smoothing out the non-analytic behavior present at $N=\infty$.
Although not shown here, the real part of $C_{yy}$ remains finite in the region 
$k\leq |\w|\leq \sqrt{4m^2+k^2}$, although non-analytic dependence appears at the 
upper boundary of that region. In fact, we find that there is a logarithmic branch cut that appears
at $\w_\star=\sqrt{(2m)^2+k^2}$, and at $-\w_\star$. 
We have previously identified\cite{ws,will} its zero-momentum manifestation in the conductivity, 
where in addition $\w_\star=2m$ becomes a zero of $\s=i\Pi^T(\w,0)/\w$.
At finite but small momentum, $k\ll T$, the branch point disperses quadratically away from its $k=0$ position, 
$\w_\star-2m=k^2/(4m)$, and eventually acquires a relativistic dispersion $\w_\star=k$, when $k\gg T$.
In terms of the real-frequency behavior, both the field theory and holography yield the same small- and 
large-$w$ asymptotics: $\Im C_{yy}\propto \w$
in both limits, in agreement with the oddness requirement and dimension of $C_{\mu\nu}$. As momentum
tends to zero, the slope of the linear part at small frequencies increases
yielding a peak of increasing height for $\Im C_{yy}/w$:
this is the formation of the delta function peak of the conductivity obtained in the $q=0$ limit.
We have indeed verified that
at sufficiently small $q$, the peak rapidly saturates the weight of the $q=0$ delta function.
Such behavior is naturally absent in the holographic analysis, where the conductivity of the
\emph{interacting} boundary CFT remains finite in the d.c.\ limit.
%We further note that the peak location moves linearly with momentum.  

We now examine the spectral density of the charge density auto-correlator, $\Im C_{tt}$, which is plotted
in \rfig{rot-Ctt}. Just as $\Im C_{yy}$, it vanishes in the kinematically forbidden region,
but now shows a jump discontinuity at the pair-production threshold, $\pm\w_\star=\pm\sqrt{(2m)^2+k^2}$.
This leads to a logarithmic divergence of the real part at $\pm\w_\star$.
The function $\Im C_{tt}(w,q)$ is of course $w$-odd, but instead of vanishing linearly at small frequencies 
as required by hydrodynamics, it diverges as $1/w$ (in the limit of $q=0$, a double instead of a single pole emerges at the origin
so that the real part of the conductivity acquires a delta function). The finite-momentum pole at $w=0$ arises because 
when $N=\infty$, diffusion
does not take place as the theory is free of interactions: the conserved charge must propagate ballistically
at all times.
%The diffusion constant vanishes and the hydrodynamic form reduces to $C_{tt}\sim 1/(iw-Dq^2)\ra 1/iw$ in accordance with \rfig{rot-Ctt}. %\subsection{Finite $N$}
At order $1/N$, interactions between the quasiparticles appear and as
a consequence so does charge diffusion. The precise value of the 
large-$N$ diffusion constant, $D=0.249 N/T$, was deduced\cite{book} from the 
Einstein relation and knowledge of the charge susceptibility\cite{book} and d.c.\ conductivity\cite{will},
which are respectively (to leading order in $1/N$): $\chi=\sqrt{5}\Theta T/2\pi$ and $\s_{\rm dc}=0.085 N$.
The divergence of the diffusion constant in the $N\ra\infty$ limit is in agreement with its
interpretation as a scattering time between the critical quasiparticles. 
In the relativistic limit at large momentum, the spectral function $\Im C_{tt}$ gains resemblance with the $T=0$
form, $1/\sqrt{-w^2+q^2}$, having essentially no weight in the region $w<q$ (except for the $w=0$ pole) 
and develops a sharp peak at $w\sim q$.  

We have seen that at $N=\infty$, the current correlators show branch cuts in the LHP
due to the absence of interactions. At finite $N$, we expect the branch points to disappear from the real
axis and the spectral weight to spread over all frequencies
in line with general expectations, and with the holographic results.
One possibility is that the branch cuts split into a discrete sequence of poles and zeros,
in analogy with the QNMs that have been discussed above. One such QNM was already found\cite{ws,will} at $\mc O(1/N)$,
the D-pole of the transverse correlator, which leads to a pole in the small-frequency
conductivity. Another possibility is that the branch cuts move away from the real axis in the LHP.

\subsection{Sum rule}
The $N=\infty$ theory satisfies the sum rule \req{sr1}, where the static correlator
on the r.h.s., $\Pi^T(0,q)$, can be numerically computed and has precisely the same form
as what was found for the holographic models:
\begin{align}\label{eq:rot-tanh}
  \Pi^T(0,q)=T\s_{\infty} q \tanh(\al q)\,,
\end{align}
where $\al\approx 0.466$, as we show in \rfig{sr-rot}. The simplified expression used to calculated
the static correlator can be found in Appendix C of Ref.~\onlinecite{ws}.
We note that $\al$ differs by less than one percent from
$T/(2\pi \chi)$, where $\chi=\sqrt{5}\Theta T/2\pi$ as given above. 
The $q=0$ version, \req{sr1-sig}, i.e.\ the sum rule for the
conductivity was shown to hold previously by us\cite{ws}. Just as in the holographic case,
we find that $-(\Im\Pi^T/w)-\s_\infty$ decays as $1/w^2$ at large frequencies
in line with the asymptotic Lorentz invariant form.
\begin{figure}
\centering
\includegraphics[scale=.5]{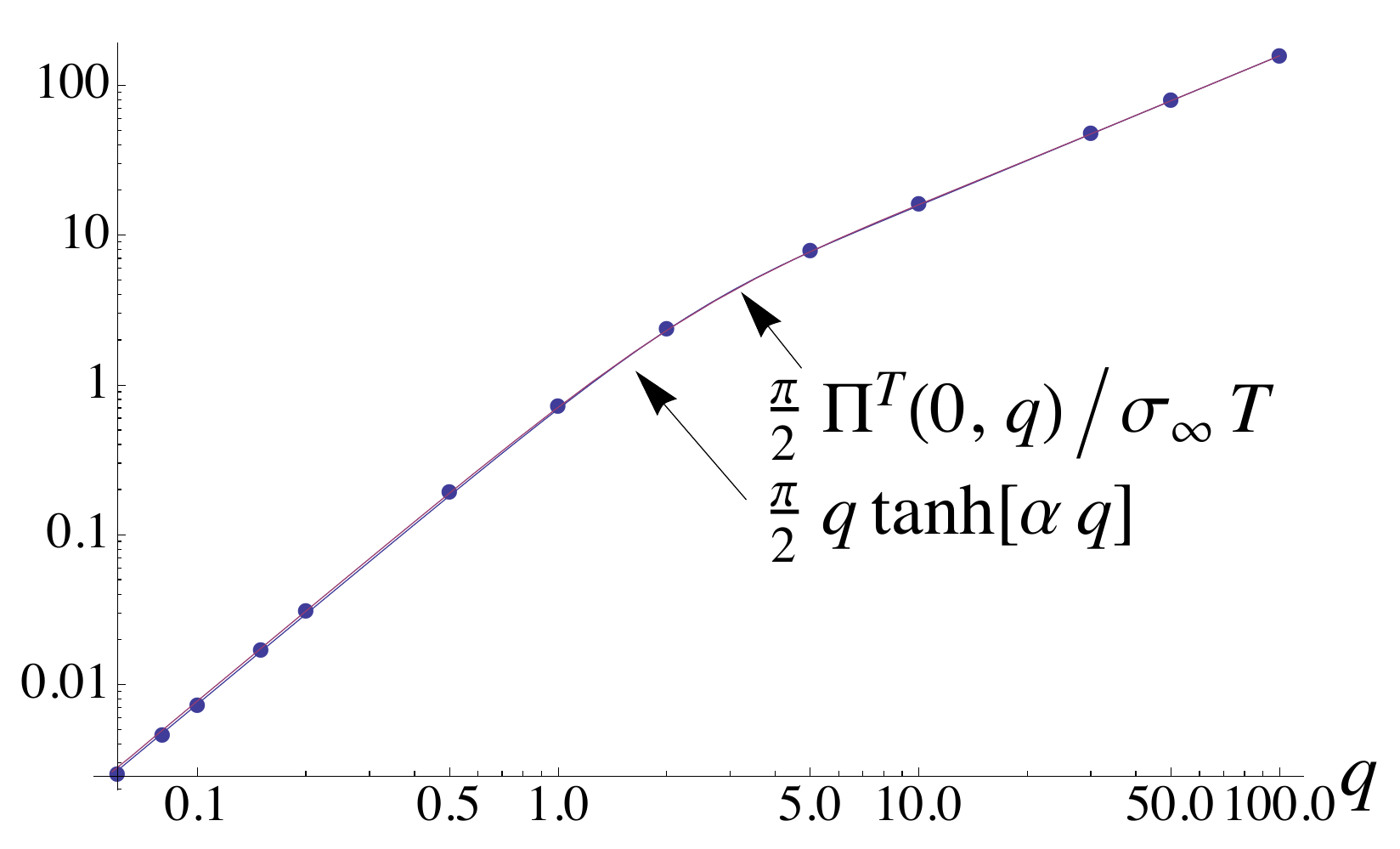}
\caption{\label{fig:sr-rot} Verification of sum rule \req{sr1} for the quantum critical vector O($N$) model
in the $N\ra \infty$ limit. Note the resemblance with the holographic result, \rfig{sr}.
}  
\end{figure}

\section{Conclusions}

This paper has described details of the current correlators of CFT3s represented holographically by the 
Einstein-Maxwell action in Eq.~(\ref{eq:S_bulk}), augmented by the Weyl term in Eq.~(\ref{eq:4-deri}). 
Such an approach has been argued \cite{myers11} to be the most general representation of the two-point
correlator unto 4 derivatives in the holographic theory. In our previous paper \cite{ws} we demonstrated that the poles
and zeros of the response function identified the quasinormal modes of the holographic theory, and led to a simple and complete
description of the frequency-dependent conductivity. These quasinormal modes replace the quasiparticle excitations of a traditional Boltzmann
analysis of quantum transport. The present paper has extended such an analysis to spatially modulated response functions, which is linked
to the dispersion of the quasinormal modes as a function of momentum. 

We refer the reader back to Section~\ref{sec:main} for a more detailed summary
of our results, and focus on the main points here. We have found that the thermal current correlators obtained
holographically bear a strong imprint of the $T=0$ Lorentz invariance, as one can see from the asymptotics
and the behavior under exchange of momentum and frequency. Crucially, their QNM spectra were seen to be 
a useful tool in understanding the nature of the excitation modes of the CFT. For instance, sharp
transitions in the QNM distribution were found to correspond to hydrodynamic-to-relativistic crossovers 
in the real-frequency response functions. In this respect, it was found that the presence of the four-derivative Weyl term
can lead to distinct behavior. The two possible cases for the crossover mechanism
are epitomized in \rfig{cross-2-cases}, where in case 1 
the hydrodynamic pole and the special purely damped D-pole of the charge correlator, $C_{tt}$, eventually detach from the imaginary axis (\rfig{zip}),
while in case 2, the hydrodynamic pole and D-\emph{zero} cannot detach and instead move down the axis.
Case 2 applies to $\g< 0$, while case 1 to $\g\geq 0$, which includes the pure two-derivative theory describing
the R-currents, where although the D-QNM is missing at zero momentum, a substitute appears for $q>0$.
It was further shown that S-duality, realized as electric-magnetic duality in the bulk, is an integral
part of the physics, even for the direct theory. We have finally discussed a number of sum rules generalizing to finite momentum 
those for conductivity\cite{sum-rules,ws}, some of them in excellent agreement with equivalent statements in the
large-$N$ vector O($N$) model. A natural connection between the sum rules and gauge invariance in the holographic bulk was made. 
It would be interesting to see if such a connection generalizes to other holographic sum rules such as those
derived in Ref.~\onlinecite{sum-rules}.

We conclude our discussion by comparing the holographic description of the 
quantum-critical conductivity to quantum Monte Carlo results
on the O(2) critical point obtained some time ago.\cite{sorensen} The O(2) model 
has \cite{sorensen,fazio} $\sigma_\infty \approx 0.33 Q^2 /h$ in the $\w/T\ra\infty$ limit, 
where $Q$ is the charge of the bosons. 
The analytic continuation of imaginary time Monte Carlo data gave
the estimate for the d.c.\ conductivity at non-zero temperature of\cite{sorensen} $\sigma (0) \approx 0.45 Q^2/h$, which yields the
ratio $\sigma(0)/\sigma_\infty \approx 1.36$. The holographic prediction of the four-derivative theory is \cite{myers11} 
$\sigma(0)/\sigma_\infty = 1 + 4 \gamma$,
along with the bound $|\gamma| \leq 1/12$. Interestingly, the maximum possible holographic value at $\gamma=1/12$ of 1.33 is very close to the current quantum Monte Carlo estimate. This value of $\gamma$ is also consistent with considerations \cite{suvrat} from $T=0$ multipoint correlators of the CFT3, when computed in a vector large-$N$ expansion of the S-dual conformal gauge theory. In future work, it would be interesting to obtain
additional Monte Carlo data which allow comparison with the frequency and momentum dependence predicted by the holographic theory.

\acknowledgments
We are grateful to S.~Hartnoll and C.P.~Herzog for their comments on the manuscript, and to
V.\ Cardoso for discussions regarding the QNM spectrum.
The research was supported by the U.S.\ National Science Foundation under grant DMR-1103860, and by 
the U.S.\ Army Research Office Award W911NF-12-1-0227 (S.S.); and by the Walter Sumner Foundation (W.W.-K.). Research at Perimeter Institute is supported by the Government of Canada through Industry Canada and by the Province of Ontario through the Ministry of Research \& Innovation.

\appendix
\section{Properties of QNMs}
\label{ap:qnm}
In this appendix we discuss various properties of the momentum-dependence of the QNMs of 
the $\g=0$ theory in more detail.
\begin{figure}[h]
\centering
\includegraphics[scale=.5]{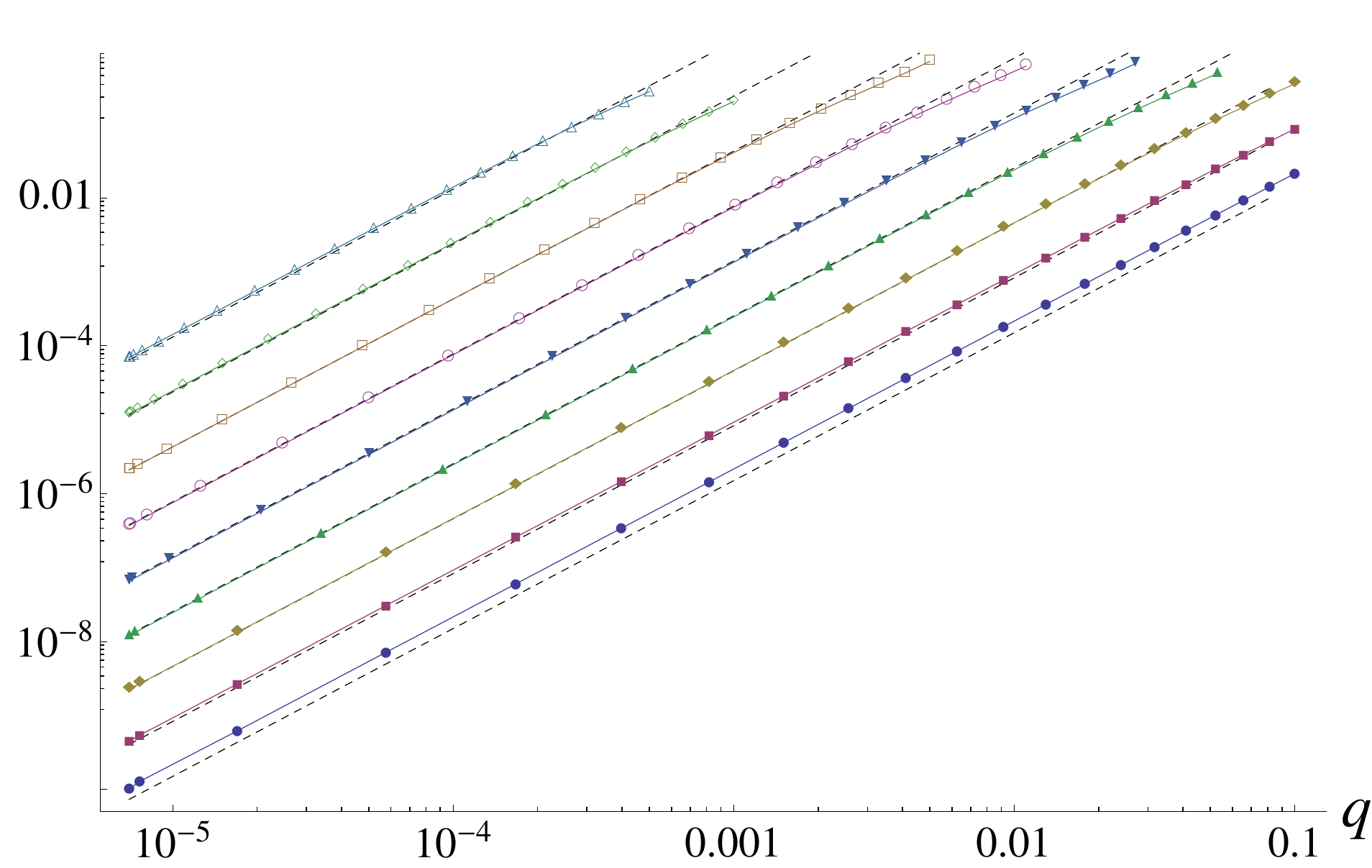}
\caption{\label{fig:all-quad} Momentum dispersion of the first 9 members of the tower of purely damped QNMs
that propagate towards $w=0$. Their
real part vanishes, and we plot the imaginary part relative to the zipping point out of which they 
emerge at $q=0$: $w_n-w_n^{\rm zip}= i\al_n q^2$. From bottom to top: $n=1$ to $9$. The dashed
lines are quadratic dispersions with the coefficients satisfying $\al_n\approx a_1 e^{a_2 n}$,
$(a_1,a_2)=(0.27,1.71)$.}  
\end{figure} 

\subsection{Quadratic dispersion}
At $\g=0$ and for infinitesimal momentum $q$, all the QNMs of the current correlators appear on the imaginary axis 
at the special zipping frequencies $w_n^{\rm zip}=-i3n/2$, where $n\geq 0$ is an integer. 
The diffusive pole of $C_{tt}$ is already present even at $q=0$. Just as the diffusive pole,
all the QNMs, i.e.\ both poles and zeros of $C_{tt}$, disperse quadratically with momentum away from 
their respective zipping point. This
is shown in \rfig{all-quad}, where we only show the QNMs propagating towards zero, $w_n=w_n^{\rm zip}+i\al_n q^2$,
with $\al_n>0$, with $1\leq n\leq 9$. The $n=1$ upward propagating mode is a pole of $C_{tt}$, $n=2$ is
a zero, and so forth. 
The partner QNMs propagating toward $-i\infty$ disperse as $w_n^{\rm zip}-i\al_n q^2$ and
are thus not shown for clarity. As discussed in the main body,
the dispersion coefficients grow exponentially with $n$. This can be seen form the dashed lines in \rfig{all-quad}.
In particular, this exponential growth of the dispersion coefficients implies that the unzipping process
will occur exponentially fast as $q$ is tuned from zero to $0.557$, at which point the last QNMs detach from the
imaginary axis. 

\subsection{Double poles and zeros}

\begin{figure}
\centering
\includegraphics[scale=.45]{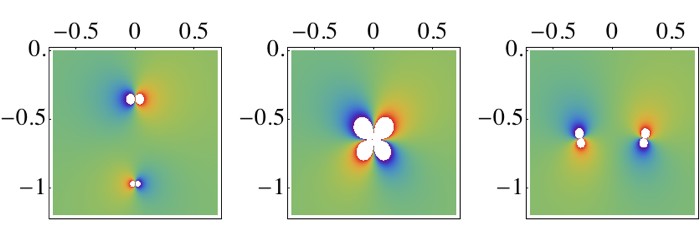}%.48   
\caption{\label{fig:double} Formation of a double pole for $C_{tt}(w,q)$ as a function of $q$
at the intermediate step of the unzipping process. Plots of $-\Im C_{tt}$ are in the LHP $w$-plane. 
Left to right: $q=0.5,0.557, 0.6$. Red represents positive values while blue negative ones.
}  
\end{figure}

In the main body, we described the generic unzipping process that occurs as a function of 
momentum. It also occurs at zero momentum as a function of $\g$ at zero momentum as noted 
previously\cite{ws}. Here we substantiate the claim by plotting the charge density spectral function
$-\Im C_{tt}(w,q)$ as a function of momentum. As $q$ changes from $q<q_c$ to $q>q_c$, a double
pole appears directly at $q_c=0.5573187$, as shown in \rfig{double}. As we are plotting the imaginary part of $C_{tt}$
not the norm, the phase structure shows very clearly that we have a double pole at $q_c$, with
four lobes, in contrast to single poles that have only two lobes.

\subsection{Lifetimes}
\begin{figure}
\centering
\includegraphics[scale=.55]{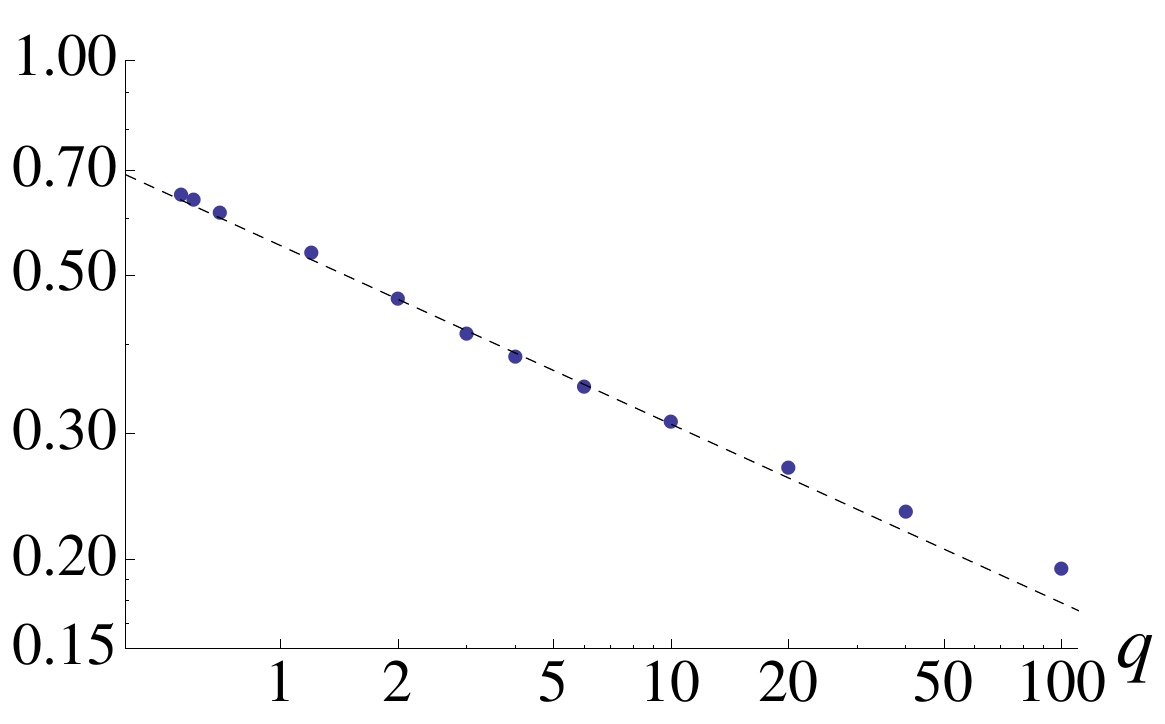}%.48   
\caption{\label{fig:inv-life} Absolute value of the imaginary part of the QNM closest to the
real frequency axis as function of momentum. The dashed line is $\propto 1/q^{1/4}$.
}  
\end{figure}

Still at $\g=0$, we now take a closer look at the rate at which the imaginary part of the QNM closest 
to the real axis approaches zero. We track the imaginary part of the QNM after it detaches from the
imaginary axis at momentum $q_c$ as described in the previous subsection. The result is shown in \rfig{inv-life},
where we see that the imaginary part vanishes relatively slowly, slower than $1/q^{1/4}$ in the 
regime we have investigated. We recall that the absolute value of the imaginary part should correspond to 
the inverse lifetime of that mode or quasiparticle, while the real part describes its excitation energy. In contrast, the real
part scales linearly with momentum. 

\section{Analysis of differential equations}
\label{sec:heun}

We discuss the underlying structure of the ODEs appearing in the main body, namely those for the
transverse gauge field $A_y$, and its EM-dual, $\hat A_y$. This sheds some light on the physics
and provides an additional mathematical handle on the problem.

\subsection{$\gamma=0$: Heun's equation}
\begin{figure}
\centering%
\subfigure[~$\g>0$]{\label{fig:rsp_gp}\includegraphics[scale=.4]{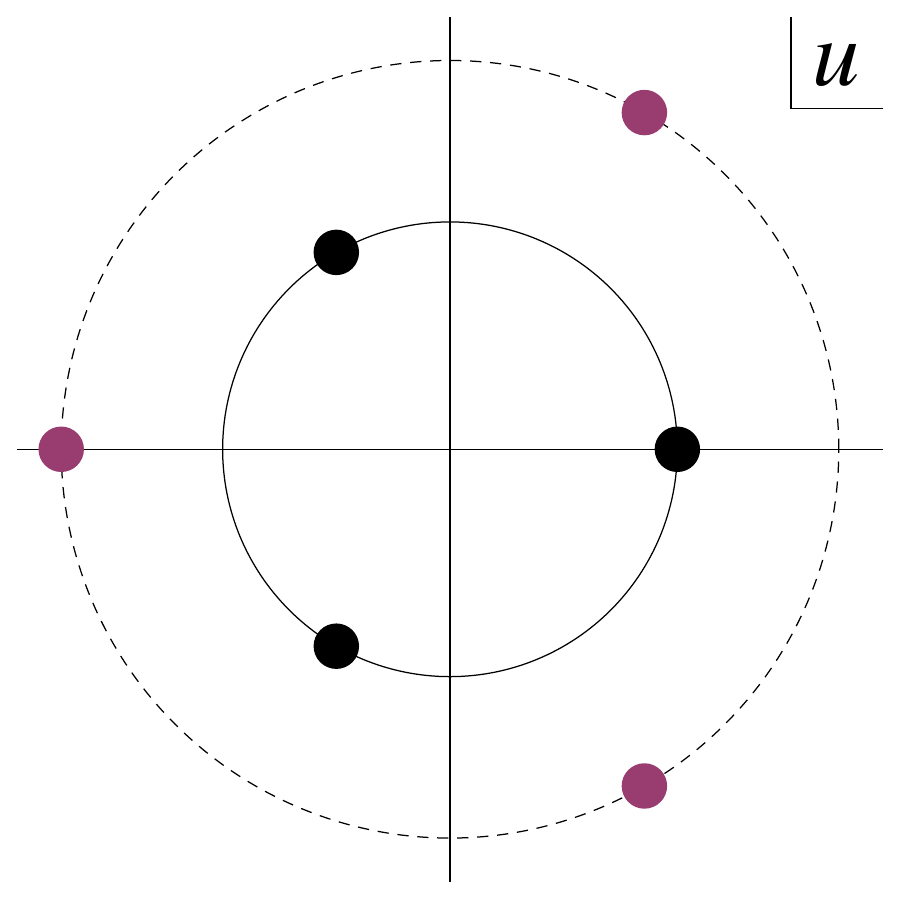}}  
 \subfigure[~$\g<0$]{\label{fig:rsp_gn} \includegraphics[scale=.4]{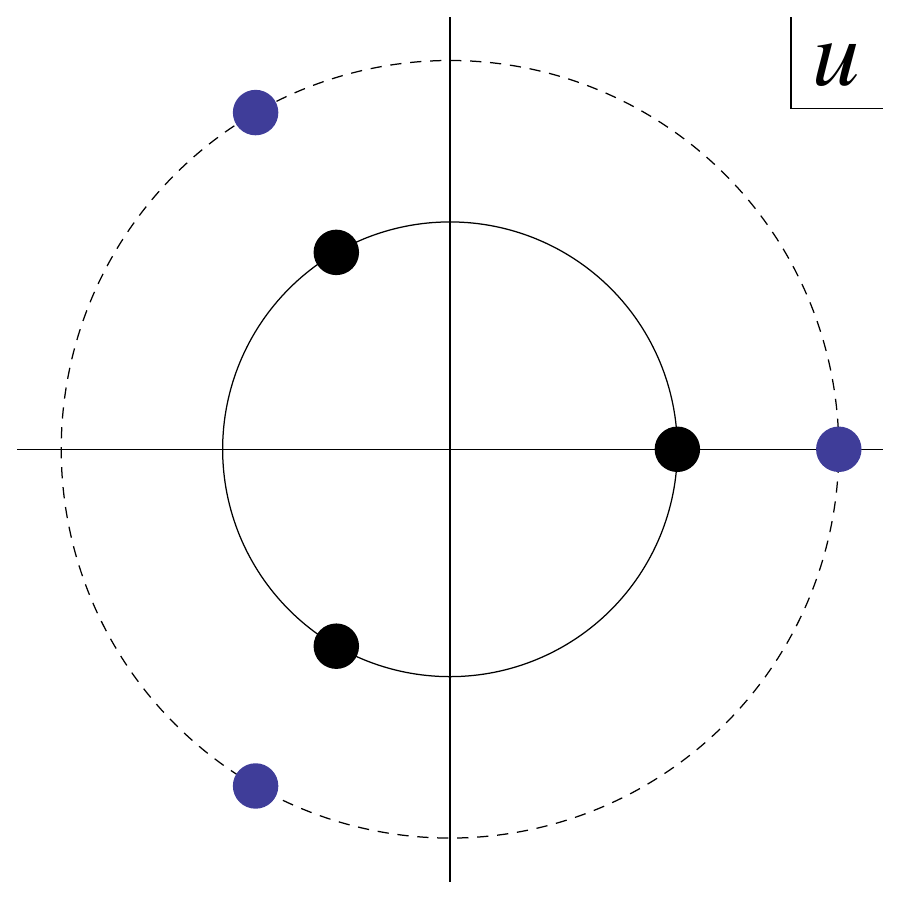}}
\caption{\label{fig:rsp} Regular singular points (RSPs), excluding the one at $\infty$, of the
EoM of the transverse gauge mode $A_y$ in the complex $u$-plane. The 3 on the inner circle 
correspond to the horizon, and its 2 complex partners.
The outer radius (dashed) has radius $1/(4|\g|)^{1/3}$. When $\g=0$, the outer-radius RSPs disappear
and only 3 finite RSPs remain.
} 
\end{figure}

Let us first discuss the $\gamma=0$ EoM for $A_y(u)$, 
\begin{align}\label{eq:ap-eomAy}
  A_y''+\frac{f'}{f}A_y' + \frac{w^2-q^2f}{f^2}A_y=0\,, \qquad f(u)=1-u^3\,.
\end{align}
It is a \emph{Fuchsian} ODE, namely it is a linear and homogeneous differential
equation with a finite number of singular points on the Riemann 
sphere $\mathbb CP^1=\mathbb C\cup \{\infty\}$, all of them
\emph{regular}. We are thus including the point at infinity in our treatment. In terms of the coordinate $u=1/r$, this
point corresponds to the black hole singularity at $r=0$. The above ODE has 4 regular singular points (RSPs):
$1,\z,\z^2=\z^* $ and $\infty$, where $\z=e^{-i2\pi/3}$ so that the first 3 RSPs are the cubic roots of unity,
the zeros of $f=1-u^3$, they are shown on the solid circle of
\rfig{rsp}. In the same way that all second order Fuchsian ODEs with 3 RSPs can be mapped to
the hypergeometric equation, all the ones with 4 RSPs can be mapped to the so-called 
\emph{Heun equation}\cite{maier,NIST:DLMF}:
\begin{align}\label{eq:heun}
  \frac{d^2H}{ds^2}+\left( \frac{\de_1}{s}+\frac{\de_2}{s-1}+\frac{\de_3}{s-a} \right)\frac{dH}{ds}
  +\frac{\al\be s- Q}{s(s-1)(s-a)}H=0
\end{align}
where $\de_3=\al+\be-\de_1-\de_2+1$. This equation has RSPs at $0,1,a,\infty$.
Our original equation can be mapped to this one via the following simple canonical transformations.
First, we perform the linear change of variables: 
\begin{align}
  u=(\z-1)s+1\,. 
\end{align}
This maps the RSPs as
follows: $1\ra 0$, $\z\ra 1$, $\z^2\ra -\z^2=:a$ and preserves the one at $\infty$.
Second, we define: 
\begin{align}\label{eq:transf2}
  A_y=s^{b_1}(s-1)^{b_2}(s-a)^{b_3}H(s)\,,
\end{align}
where the exponents $b_1,b_2,b_3$ are chosen
to remove the terms proportional to $1/s^2$, $1/(s-1)^2$ and $1/(s-a)^2$, respectively, 
that appear in the transformed equation. The algebraic equations for the $b_i$ are readily found to be:
\begin{align}\label{eq:indicial}
  b_i^2+\frac{w^2}{9}u_i=0\,,\quad i=1,2,3\,,
\end{align}
where $(u_1,u_2,u_3)=(1,\z,\z^2)$ are the RSPs excluding $\infty$. These are precisely the
indicial equations of the corresponding RSPs of the original ODE \req{ap-eomAy}; their solutions yield the 2 characteristic 
exponents associated with each RSP. This is not so surprising as the indicial equation at $u_i$ can be obtained
by substituting the power law solution $(u-u_i)^{b_i}$ in the ODE \req{ap-eomAy} and solving for $b_i$ in the limit where $u\ra u_i$.
The solutions of \req{indicial} are $b_i=\pm iwu_i/3$. We note that \req{indicial} is not the indicial equation in the
new variable $s$, as the latter equation is transformed and the corresponding characteristic exponents differ.
The transformation \req{transf2} shifts both characteristic exponents at each RSP $u_i$ by $-b_i$, and the ones
associated with the RSP at $\infty$ by $b_1+b_2+b_3=0$. 

Fixing the $b_i$ as we have just described yields the desired form; it only remains to read off the Heun parameters.
The resulting Heun ODE (and its solutions) are determined by the so-called Riemann P-symbol
and the \emph{accessory} parameter $Q$\cite{maier}: 
\begin{align}\label{eq:Psymb}
  P &\begin{Bmatrix} 0 & 1 & a & \infty &  \\
  0 & 0 & 0 & \al & ;\, s \\
   1-\de_1 & 1-\de_2 & 1-\de_3 & \be &
 \end{Bmatrix} = 
  P\begin{Bmatrix} 0 & 1 & -\z^2 & \infty & \\
  0 & 0 & 0 & 0 & ;\, s \\
   2i\frac{w}{3} & 2i\frac{w}{3}\z & 2i\frac{w}{3}\z^2 & 2 &
 \end{Bmatrix} \\ \nn\\
 Q &=iq^2\z/\sqrt{3}\,,
\end{align}
where $\z=e^{-i2\pi/3}$ as defined above. 
The first row of the Riemann P-symbol
contains the RSPs and the other 2 rows the corresponding characteristic exponents. Since the
$b_i$'s determine the characteristic exponents: $\de_i=1+2b_i=1+2(\pm iwu_i/3)$, we have the freedom
to choose the sign of each $b_i$. We opt for the solutions $b_i=-iwu_i/3$, motivating our choice by 
the in-falling boundary condition at the horizon used in the main body, which requires $A_y\sim (1-u)^b$ near $u=1$ 
with $b=-iw/3$. Thus the choice $b_1=b$. 

We see that the frequency $w$ is the ``central'' parameter, since it determines 3 of the
4 non-zero characteristic exponents, the exception being the one at $\infty$. The momentum $q$ enters only via
the accessory parameter $Q$. 
\subsubsection{Solution}
There are $8=2\times 4$ local solutions to the Heun equation\cite{NIST:DLMF,maier}: 2 per RSP. The one corresponding
to the RSP $s=0$ and the characteristic exponent $0$ is called the \emph{local Heun function} 
and denoted by $H\!\ell(a,Q;\al,\be,\de_1,\de_2;s)$. It is normalized to $1$ at $s=0$.
Incidentally, amongst the $8$ local solutions, this is the one that is relevant for us since 
the $s=0$ RSP corresponds to the black hole 
horizon at $u=1$, and this is precisely where we apply the in-falling boundary condition. 
Interestingly, it is known that when $\de_1$ is a non-positive integer, $H\!\ell$ is ill-defined\cite{maier}. Since
$\de_1=1-2iw/3$, this actually occurs at the special zipping frequencies identified in the bulk, $w_n^{\rm zip}=-i3n/2$, 
$n\in \mathbb Z^+$. As we discuss below, these correspond to poles of $H\!\ell$.
The solution of \req{ap-eomAy} can thus be expressed using
$H\!\ell$:
\begin{align}
  %F &:=\frac{A_y}{(1-u)^b}= 
A_y &= (1-\z)^{-iw/3}s^{-iw/3}(1-s)^{-iw\z/3}(1-s/a)^{-iw\z^2/3}  \nn\\
 &\qquad \times H\!\ell \left(-\z^2,i\frac{q^2}{\sqrt{3}};0,2,1-2i\frac{w}{3},1-2i\frac{w\z}{3};s \right) 
 \label{eq:Ay-heun}\\
  s &=\frac{u-1}{\z-1}
\end{align}
where at the horizon, $s=0=u-1$, $F=A_y/(1-u)^{-iw/3}$ is normalized to unity. Note the appearance of the
variable $u$ in the normalization condition instead of $s$. %This ratio was defined as $F(u)$ in previous works.

The local Heun function has a Fuchs-Frobenius series expansion in the disk $|s|<1$, \\
$H\!\ell(a,Q;\al,\be,\de_1,\de_2;s)=\sum_{j=0}^\infty c_j s^j$, with the coefficients satisfying\cite{NIST:DLMF}
\begin{align}
  c_0=1\,,\quad c_1 a\de_1-c_0 Q=0 \\
  R_j c_{j+1}-(Q_j+Q)c_j+P_j c_{j-1}=0
\end{align}
The parameters $P_j,Q_j,R_j$ used in the recursion relations are defined as
\begin{align}
  P_j &=(j-1+\al)(j-1+\be) \\
  Q_j &= j[(j-1+\de_1)(1+a)+a\de_2+\de_3] \\
  R_j &=a(j+1)(j+\de_1)
\end{align}
Contrary to its simpler hypergeometric cousins, no closed-form solution for the $c_j$ 
is known in general. Nonetheless, it can be efficiently implemented
and we have explicitly verified that the series solution matches the one obtained 
by numerically solving the initial-value problem. 

Using \req{dict-PiT}, the solution for the transverse correlator $C_{yy}(w,q)=\Pi^T(w,q)$ can be thus
expressed as
\begin{align}\label{eq:PiT-heun}
  \frac{\Pi^T(w,q)}{-\chi_0}= iw + \frac{e^{i5\pi/6}}{\sqrt 3}
\frac{\pd_s H\!\ell(-\z^2,i\frac{q^2}{\sqrt{3}};0,2,1-2i\frac{w}{3},1-2i\frac{w\z}{3};s_0)}{\;\;\,H\!\ell(-\z^2,i\frac{q^2}{\sqrt{3}};0,2,1-2i\frac{w}{3},1-2i\frac{w\z}{3};s_0)}
\end{align}
where $s_0=1/(1-\z)=-e^{i5\pi/6}/\sqrt 3$ is the value of $s$ at $u=0$.

In special cases, the Heun function can be simplified. For instance,
when $\al\be=Q=0$, the Heun equation looses its RSP at $s=\infty$ and is said to be trivial.
In our case, this happens when the momentum vanishes, $q=0$. It was previously found\cite{m2cft} that
the $q=0$ EoM for $A_y$ can be analytically solved: $A_y=e^{iwz/3}$, where the so-called
tortoise coordinate reads
$z(u)=-\ln(1-u)-\z\ln(1-u/\z)-\z^2\ln(1-u/\z^2)$ (see Ref.~\onlinecite{ws} for instance). Equivalently,
$A_y=(1-u)^{b_1}(1-u/\z)^{b_2}(1-u/\z^2)^{b_3}$ and the Heun function $H\!\ell$ is
simply a constant, as can be seen from \req{Ay-heun}, or directly from the recursion relation
given above. In that case, using \req{PiT-heun} we recover $\Pi^T(w,q)=-i\chi_0 w$ since $\pd_sH\!\ell=0$.
In some other cases, the Heun equation also looses a singularity and
can be mapped to the hypergeometric equation\cite{maier}, for e.g.\ when $\de_3=0$, $Q=\al\be a$
or when $Q=\de_1=0$. For our ODE, this always implies $q=0$ and is thus of no interest.

We make a remark regarding the QNMs of $\Pi^T=-\chi_0 A_y'(0)/A_y(0)$. As remarked above,
$H\!\ell$ in \req{PiT-heun} has poles
directly at the zipping frequencies $w_n^{\rm zip}=-i3n/2$, $n\geq 1$. These correspond to
poles of $A_y(0)$ but not to zeros of $\Pi^T$ since they are canceled
by the poles of $A_y'(0)$, which are given by the poles of the derivative of $H\!\ell$. 
Thus, only the zeros of $A_y'(0)$, i.e.\ those of $H\!\ell$, give the QNM poles of $\Pi^T$.

\subsection{$\g\neq 0$: more singular points}
We now consider the $A_y$-EoM including the 4-derivative term with coupling $\gamma$:
\begin{align}\label{eq:ap-eomAy-g}
  A_y''+\left(\frac{f'}{f}+\frac{g'}{g}\right)A_y' + \frac{w^2-q^2f(1-8\g u^3)/g}{f^2}A_y=0\,, \qquad f(u)=1-u^3\,.
\end{align}
The main difference with the $\g=0$ equation is the appearance of the term $g'/g$, where 
$g=1+4\g u^3$. The equation is again Fuchsian, but when $\g$ is finite and 
satisfies $|\g|<1/12$, it has 3 additional RSPs compared with \req{ap-eomAy}, for
a total of 7 RSPs. The new RSPs are the zeros of $g(u)$: 
\begin{align}
  -\frac{\sgn(\g)}{(4|\g|)^{1/3}}(1,\z,\z^2)=:(v_1,v_2,v_3)
\end{align}
Thus, they differ from the 3 at $\g=0$ by a factor of $-\sgn(\g)/|\g|^{1/3}$. The RSP at $\infty$ remains.
The 6 finite RSPs, $\{u_i,v_i\}$, are shown in \rfig{rsp} for both signs of $\g$.

The additional 3 RSP found at finite $\g$, however, play a different role from the original 3 finite ones,
$\{u_i\}_{i=1}^{3}=\{1,\z,\z^2\}$. This is expected on physical grounds: the latter are the horizons of
the black hole ($u=1$ is the real horizon, while the other are the ``complex horizons'') and as such
lead to singularities in the metric. The new ones gained at finite $\g$ have nothing to do with the metric
(the gauge field is in the probe limit) and they only affect the EoM of the gauge field. This can be seen more mathematically
as well: \emph{the characteristic exponents of the $v_i$ RSPs vanish identically.} 
Moreover, their presence does not affect 
those of the $u_i$, which were found to be $\pm iw u_i/3$. One of the characteristic exponents of the RSP
at $\infty$ changes however from $2$ to $5$, as must be the case because of Fuch's relation: 
$\sum_\al (\rho_1(\al)+\rho_2(\al)-1)=-2$, where $\al$ sums over all RSPs and the $\rho_i$ are the 2 exponents
associated with each point. 

Finally, we note that the EoM for the S-dual gauge field $\hat A_y$, \req{eomAy-g-dual}, has 3 additional
RSPs, for a total of 10 RSPs. The new ones are the zeros of $1-8\g u^3$: 
$(\hat v_1,\hat v_2,\hat v_3)=\tfrac{\sgn(\g)}{(8|\g|)^{1/3}}(u_1,u_2,u_3)$. Just like the $v_i$, these also have frequency
and momentum independent characteristic exponents, in this case $\{0,1\}$ instead of $\{0,0\}$ for the $v_i$. From
Fuch's relation (mentioned in the previous paragraph), we see that the presence of these RSPs does not
alter the characteristic exponents of the RSP at $\infty$, making them even more innocuous than the $v_i$.
Indeed, we note that they have the same characteristic exponents as a \emph{regular} point.

\bibliography{ads-q}{}
\end{document}